\documentclass[journal,draftcls,onecolumn,10pt,twoside]{IEEEtran}

\usepackage{cite}
\usepackage{graphicx}

\newcommand{\uge}{\mathbin{\rotatebox[origin=c]{90}{$\ge$}}}

\usepackage[cmex10]{amsmath}

\usepackage[justification=centering]{caption}
\usepackage{float}
\usepackage{verbatim}
\usepackage{relsize}
\usepackage{amsfonts}
\usepackage{amsthm}
\usepackage{amssymb}
\usepackage{latexsym}
\usepackage{url}
\usepackage{epstopdf}
\usepackage{enumerate}
\usepackage{mathrsfs}
\usepackage{tabulary}
\usepackage{color}
\usepackage{balance}
\newtheorem{theorem}{Theorem}
\newtheorem{Lemma}{Lemma}

\newtheorem{proposition}{Proposition}

\newcommand{\floor}[1]{\left\lfloor #1\right\rfloor}

\begin{document}
\title{Bounds on the Size and Asymptotic Rate of Subblock-Constrained Codes}

\author{Anshoo~Tandon,~\IEEEmembership{~Member,~IEEE},~Han~Mao~Kiah,~\IEEEmembership{~Member,~IEEE},\\ Mehul~Motani,~\IEEEmembership{Senior Member,~IEEE}
	\thanks{A. Tandon and M. Motani are with the Department of Electrical and Computer Engineering, National University of Singapore, Singapore 117583 (email: anshoo@nus.edu.sg, motani@nus.edu.sg).}%
    \thanks{H.~M.~Kiah is with the School of Physical and Mathematical Sciences, Nanyang Technological University, Singapore 637371 (email: hmkiah@ntu.edu.sg).}%
    }%

\maketitle

\begin{abstract}
The study of subblock-constrained codes has recently gained attention due to their application in diverse fields. We present bounds on the size and asymptotic rate for two classes of subblock-constrained codes. The first class is binary \emph{constant subblock-composition codes} (CSCCs), where each codeword is partitioned into equal sized subblocks, and every subblock has the same fixed weight. The second class is binary \emph{subblock energy-constrained codes} (SECCs), where the weight of every subblock exceeds a given threshold. We present novel upper and lower bounds on the code sizes and asymptotic rates for binary CSCCs and SECCs. For a fixed subblock length and small relative distance, we show that the asymptotic rate for CSCCs (resp.~SECCs) is strictly lower than the corresponding rate for constant weight codes (CWCs) (resp.~heavy weight codes (HWCs)).  Further, for codes with high weight and low relative distance, we show that the asymptotic rates for CSCCs is strictly lower than that of SECCs, which contrasts that the asymptotic rate for CWCs is equal to that of HWCs.  We also provide a correction to an earlier result by Chee et al. (2014) on the asymptotic CSCC rate. Additionally, we present several numerical examples comparing the rates for CSCCs and SECCs with those for constant weight codes and heavy weight codes. 
\end{abstract}

\vspace{5mm}

\section{Introduction}
The study of subblock-constrained codes has recently gained attention as they are suitable candidates for varied applications such as simultaneous energy and information transfer~\cite{Tandon16_CSCC_TIT}, powerline communications~\cite{Chee13}, and design of low-cost authentication methods~\cite{Chee14_MCWC}. A special class of subblock-constrained codes are codes where each codeword is partitioned into equal sized subblocks, and every subblock has the same fixed composition. Such codes were called \emph{constant subblock-composition codes} (CSCCs) in~\cite{Tandon16_CSCC_TIT}, and were labeled as multiply constant-weight codes (MCWC) in~\cite{Chee14_MCWC}.

\emph{Subblock energy-constrained codes} (SECCs) were proposed in~\cite{Tandon16_CSCC_TIT} for providing real-time energy and information transfer from a powered transmitter to an energy harvesting receiver. For binary alphabet, SECCs are characterized by the property that the weight of every subblock exceeds a given threshold. The CSCC and SECC capacities, and computable bounds, were presented in~\cite{Tandon16_CSCC_TIT} for discrete memoryless channels. In this paper, we study bounds on the size and asymptotic rate for binary CSCCs and SECCs with given error correction capability, i.e., minimum distance of the code.

\subsection{Notation}

The input alphabet is denoted $\mathcal{X}$ which comprises of $q$ symbols. An $n$-length, $q$-ary \emph{code} $\mathscr{C}$ over $\mathcal{X}$ is a subset of $\mathcal{X}^n$. The elements of $\mathscr{C}$ are called \emph{codewords} and $\mathscr{C}$ is said to have \emph{distance} $d$ if the \emph{Hamming distance} between any two distinct codewords is at least $d$. A $q$-ary code of length $n$ and distance $d$ is called an $(n,d)_q$-code, and the largest size of an $(n,d)_q$-code is denoted by $A_q(n,d)$. For binary alphabet ($q=2$), an $(n,d)_2$-code is just called an $(n,d)$-code, and the largest size for this code is simply denoted $A(n,d)$.

A \emph{constant weight code} (CWC) with parameter $w$ is a binary code where each codeword has weight exactly $w$. 
We denote a CWC with weight parameter $w$, blocklength $n$, and distance $d$ by $(n,d,w)$-CWC, and 
denote its maximum possible size by $A(n,d,w)$. 
A \emph{heavy weight code} (HWC)~\cite{Cohen10} with parameter $w$ is a binary code where 
each codeword has weight \emph{at least} $w$. 
We denote a HWC with weight parameter $w$, blocklength $n$, and distance $d$ by $(n,d,w)$-HWC, 
and denote its maximum possible size by $H(n,d,w)$. 
Since an $(n,d,w)$-CWC is an $(n,d,w)$-HWC, we have that $A(n,d,w)\le H(n,d,w)$.

A subblock-constrained code is a code where each codeword is divided into subblocks of equal length, 
and each subblock satisfies a fixed set of constraints. 
For a subblock-constrained code, we denote the codeword length by $n$, the subblock length by $L$, and 
and the number of subblocks in a codeword by $m$. For the binary alphabet $\mathcal{X} = \{0,1\}$, a CSCC is characterized by the property that each subblock in every codeword has the same \emph{weight}, i.e. each subblock has the same number of ones. A binary CSCC with codeword length $n=mL$, subblock length $L$, distance $d$, and weight $w_s$ per subblock is called an $(m,L,d,w_s)$-CSCC. We denote the maximum possible size of $(m,L,d,w_s)$-CSCC by $C(m,L,d,w_s)$. Since an $(m,L,d,w_s)$-CSCC is an $(mL,d,mw_s)$-CWC, we have that $C(m,L,d,w_s)\le A(mL,d,mw_s)$.

For providing regular energy content in a codeword for the application of simultaneous energy and information transfer from a powered transmitter to an energy harvesting receiver, the use of CSCCs was proposed in~\cite{Tandon16_CSCC_TIT}. When \emph{on-off keying} is employed, with bit-1 (bit-0) represented by the presence (absence) of a high energy signal, regular energy content in a CSCC codeword can be ensured by appropriately choosing the weight $w_s$ per subblock. A natural extension of binary CSCCs are  binary SECCs, which allow the weight of each subblock to exceed $w_s$, thereby ensuring that the energy content within every subblock duration is sufficient~\cite{Tandon16_CSCC_TIT}. A binary SECC with codeword length $n=mL$, subblock length $L$, distance $d$, and weight at least $w_s$ per subblock is called an $(m,L,d,w_s)$-SECC. We denote the maximum possible size of an $(m,L,d,w_s)$-SECC by $S(m,L,d,w_s)$. 
Since an $(m,L,d,w_s)$-SECC is an $(mL,d,mw_s)$-HWC, we have that $S(m,L,d,w_s)\le H(mL,d,mw_s)$.
Also, since an $(m,L,d,w_s)$-CSCC is an $(m,L,d,w_s)$-SECC, we have that $C(m,L,d,w_s)\le S(m,L,d,w_s)$.

The relation among code sizes is summarized below. For all $m$, and $1\le d, w_s\le L$, we have
\begin{equation}
\label{eq:CodeSize_Relation}
\begin{array}{ccc}
S(m,L,d,w_s) & \ge & C(m,L,d,w_s)\\
\uge &&\uge \\
H(mL,d,mw_s) & \ge & A(mL,d,mw_s)
\end{array}
\end{equation}

We also analyze bounds on the rate in the asymptotic setting where the number of subblocks $m$ tends to infinity, $d$ scales linearly with $m$, but $L$ and $w_s$ are fixed. In the following, the base for $\log$ is assumed to be 2. Formally,  for fixed $0 < \delta < 1$, the asymptotic rates for CSCCs and SECCs with fixed subblock length $L$, 
subblock weight parameter $w_s$, number of subblocks in a codeword $m \to \infty$, and distance $d$ scaling as $d= \lfloor mL\delta \rfloor$ are defined as
\begin{align}
\gamma(L,\delta,w_s/L) &\triangleq \limsup_{m\to \infty} \frac{\log C(m,L, \lfloor mL\delta \rfloor, w_s)}{mL} ,\label{eq:CSCC_RateDef}\\
\sigma(L,\delta,w_s/L) &\triangleq \limsup_{m\to \infty} \frac{\log S(m,L, \lfloor mL\delta \rfloor, w_s)}{mL}. \label{eq:SECC_RateDef}
\end{align}

These rate can be compared with related exponents:
\begin{align}
\alpha(\delta) &\triangleq \limsup_{n \to \infty} \frac{\log A\left(n,\lfloor n \delta  \rfloor\right)}{n}, 
\label{eq:Code_RateDef}\\
\alpha(\delta,w_s/L) &\triangleq \limsup_{n \to \infty} \frac{\log A\left(n,\lfloor n \delta \rfloor,\lfloor n w_s/L\rfloor\right)}{n},
\label{eq:CWC_RateDef}\\
\eta(\delta,w_s/L) &\triangleq \limsup_{n \to \infty} \frac{\log H(n,\lfloor n \delta \rfloor,\lfloor n w_s/L\rfloor)}{n} . \label{eq:HWC_RateDef}
\end{align}

The relation between asymptotic rates can be obtained by using the relation among code sizes in \eqref{eq:CodeSize_Relation}, and the above rate definitions. For all $1 \le w_s\le L$ and $0\le \delta \le 1$, we have
\begin{equation}
\label{eq:CodeRate_Relation}
\begin{array}{ccc}
\sigma(L,\delta,w_s/L) & \overset{\text{(b)}}\ge & \gamma(L,\delta,w_s/L)\\
\text{\scriptsize(c)}~\uge & & \text{\scriptsize (d)}~\uge \\
\eta(\delta,w_s/L) & \overset{\text{(a)}}{\ge} & \alpha(\delta,w_s/L)
\end{array}
\end{equation}

We summarize our notation on code size and asymptotic rates for CWCs, HWCs, CSCCs and SECCs in Table \ref{Table:Variables}.

\begin{table}[H]
	\centering
	\begin{tabulary}{.95\textwidth}{|C|L|}
		\hline
		\multicolumn{1}{|c|}{\textbf {Notation}} 
		&\multicolumn{1}{c|}{\textbf {Description}} \\
		\hline
		\hline
		$L$ & Subblock length \\
		$m$ & Number of subblocks in a codeword \\
		$n$ & Codeword length ($n=mL$) \\
		$d$ & Minimum distance of the code \\
		$\delta$ & Relative distance of the code ($\delta = d/n$) \\
		$w_s$ & Weight per subblock \\
		$w$ & Weight per codeword ($w = m w_s$) \\
		$\omega$ & Fraction of ones in a codeword ($\omega = w/n$) \\
		$h(\cdot)$ & Binary entropy function \\
		\hline
		$(n,d)_q$-code & General $q$-ary code \\
		$A_q(n,d)$ & Maximum size of $(n,d)_q$-code \\
		$(n,d)$-code & General \emph{binary} code \\
		$A(n,d)$ & Maximum size of $(n,d)$-code \\
		$\alpha(\delta)$ & Asymptotic rate of $(n,d)$-code \\
		\hline
		$(n,d,w)$-CWC & Constant weight code (each codeword has weight $w$)\\
		$A(n,d,w)$ & Maximum size of $(n,d,w)$-CWC  \\
		$\alpha(\delta,\omega)$ & Asymptotic rate of $(n,d,w)$-CWC \\
		$\alpha_{GV}(\delta,\omega)$ & Lower bound on $\alpha(\delta,\omega)$ using Gilbert Varshamov bound for $(n,d,w)$-CWC \\
		$\alpha_{SP}(\delta,\omega)$ & Upper bound on $\alpha(\delta,\omega)$ using sphere packing bound for $(n,d,w)$-CWC \\
		\hline	
		$(n,d,w)$-HWC & Heavy weight code (each codeword has weight \emph{at least} $w$) \\
		$H(n,d,w)$ & Maximum size of $(n,d,w)$-HWC  \\
		$\eta(\delta,\omega)$ & Asymptotic rate of $(n,d,w)$-HWC \\
		\hline
		$(m,L,d,w_s)$-CSCC & Binary constant subblock-composition code (each subblock has weight $w_s$)\\
		$C(m,L,d,w_s)$ & Maximum size of $(m,L,d,w_s)$-CSCC \\
		$\mathcal{C}(m,L,w_s)$ & Space of all CSCC words \\
		$\gamma(L,\delta,w_s/L)$ & Asymptotic rate of $(m,L,d,w_s)$-CSCC  \\
		$\gamma_{GV}(L,\delta,w_s/L)$ & Lower bound on $\gamma(L,\delta,w_s/L)$ using Gilbert Varshamov bound for $(m,L,d,w_s)$-CSCC \\
		$\gamma_{SP}(L,\delta,w_s/L)$ & Upper bound on $\gamma(L,\delta,w_s/L)$ using sphere-packing bound for $(m,L,d,w_s)$-CSCC \\
		\hline		
		$(m,L,d,w_s)$-SECC & Binary subblock energy-constrained code (each subblock has weight \emph{at least} $w_s$)\\
		$S(m,L,d,w_s)$ & Maximum size of $(m,L,d,w_s)$-SECCC \\
		$\mathcal{S}(m,L,w_s)$ & Space of all SECC words \\	
		$\sigma(L,\delta,w_s/L)$ & Asymptotic rate of $(m,L,d,w_s)$-SECC  \\
		$\sigma_{GV}(L,\delta,w_s/L)$ & Lower bound on $\sigma(L,\delta,w_s/L)$ using Gilbert Varshamov bound for $(m,L,d,w_s)$-SECC \\
		$\sigma_{SP}(L,\delta,w_s/L)$ & Upper bound on $\sigma(L,\delta,w_s/L)$ using sphere-packing bound for $(m,L,d,w_s)$-SECC \\
		\hline
		$G_{\alpha - \gamma}(L,\delta,w_s/L)$ & Asymptotic rate gap between CWC and CSCC, $\alpha(\delta,w_s/L) - \gamma(L,\delta,w_s/L)$ \\
		$G_{\alpha - \gamma}^{LB}(L,\delta,w_s/L)$ & Lower bound on the asymptotic rate gap between CWC and CSCC \\		
		$G_{\eta - \sigma}(L,\delta,w_s/L)$ &  Asymptotic rate gap between HWC and SECC, $\eta(\delta,w_s/L) - \sigma(L,\delta,w_s/L)$ \\
		$G_{\eta - \sigma}^{LB}(L,\delta,w_s/L)$ & Lower bound on the asymptotic rate gap between HWC and SECC \\
		$G_{\sigma - \gamma}(L,\delta,w_s/L)$ & Asymptotic rate gap between SECC and CSCC, $\sigma(\delta,w_s/L) - \gamma(L,\delta,w_s/L)$ \\
		$G_{\sigma - \gamma}^{LB}(L,\delta,w_s/L)$ & Lower bound on the asymptotic rate gap between SECC and CSCC \\
		\hline
	\end{tabulary}
	\vspace{1mm}
	\caption{Table of Notation}
	\label{Table:Variables}
	\vspace{-5mm}
\end{table}

\subsection{Previous Work}
Among the codes discussed above, although CWCs have been widely studied, the exact characterization of $\alpha(\delta,\omega)$, for $0 < \omega < 1$, has remained elusive. A good upper bound for $\alpha(\delta,\omega)$ was given in \cite{McEliece77}, by using a linear programming bound for the CWC code size. The class of HWCs was introduced by Cohen {\em et al.} \cite{Cohen10}, motivated by certain asynchronous communication problems. The asymptotic rates for HWCs was later established by Bachoc {\em et al.} \cite{Bachoc11}. 
\begin{theorem}[Bachoc {\em et al.}\cite{Bachoc11}]\label{thm:Bachoc}
Let $0\le \delta, \omega\le 1$. Then 
\begin{equation}
\eta(\delta,\omega)=
\begin{cases}
\alpha(\delta), & \text{when } 0\le \omega\le 1/2,\\
\alpha(\delta,\omega), & \text{when } 1/2\le \omega\le 1.
\end{cases} \label{eq:eta_eq_alpha}
\end{equation}
\end{theorem}

If view of the above theorem, the inequality (a) in \eqref{eq:CodeRate_Relation} is in fact an equality for $L/2 \le w_s \le L$. 

Chee {\em et al.} \cite{Chee14_MCWC} introduced the class of CSCCs and 
provided rudimentary bounds for $C(m,L,d,w_s)$. Later, constructions of CSCCs were proposed by various authors \cite{Chee14arxiv,Wang16}. The asymptotic rate for CSCCs was also studied in \cite{Chee14_MCWC}. However, an inconsistent asymptotic rate definition in \cite{Chee14_MCWC} led to an erroneous claim regarding the CSCC rate (see \cite[Prop.~6.1]{Chee14_MCWC}). In this paper, we also provide a correct statement for the CSCC rate in the scenario where the subblock length tends to infinity via Proposition~\ref{prop:Chee} in Section~\ref{sec:rates}. 

SECCs were proposed in \cite{Tandon16_ISIT}, owing to their natural application in real-time simultaneous energy and information transfer. As shown in Section~\ref{Sec:SECC_CodeSize}, the SECC space, comprising of words where each subblock has weight exceeding a given threshold, has an interesting property that different balls of same radius may have different sizes. The lower bound on the code size for such spaces, where balls of same radius may have different sizes, was studied in \cite{Gu93}, where a \emph{generalized Gilbert-Varshamov bound} was presented. The \emph{generalized sphere-packing bound}, providing an upper bound on the code size in such spaces, has been recently presented in \cite{Fazeli15,Cullina16}, using graph-based techniques.


\subsection{Our Contributions}
The contributions of this paper are as follows:
\begin{enumerate}
\item By studying the space of CSCC and SECC codewords, we compute both upper and lower bounds for the optimal CSCC code size $C(m,L,d,w_s)$ and the optimal SECC code size $S(m,L,d,w_s)$ in Section \ref{sec:bounds}.
\item We analyze the limiting behavior of ball sizes for these spaces in high dimensions, to derive both upper and lower bounds on the asymptotic rates for CSCC and SECC in Section \ref{sec:rates}.
\item For fixed $L$ and $w_s$, we demonstrate the existence of an ${\delta}_L$ such that the inequalities (b), (c), and (d) in \eqref{eq:CodeRate_Relation} are \emph{strict} for all $\delta < {\delta}_L$ (refer  Section \ref{sec:penalty}). This result implies the following: (i) Relative to codeword-based weight constraint for CWCs (resp. HWCs), the stricter subblock-based weight constraint for CSCCs (resp. SECCs), lead to a rate penalty. (ii) Relative to CSCCs, higher rates are provided by SECCs due to greater flexibility in choosing bits within each subblock (in contrast to Theorem~\ref{thm:Bachoc}).
\item We quantify the rate penalty due to subblock-based constraints in Section \ref{sec:numerical}, by numerically evaluating the corresponding rate bounds. 
\item We also provide a correction to a result by Chee {\em et al.} \cite{Chee14_MCWC}, on the asymptotic CSCC rate in the scenario where the subblock length tends to infinity (see Proposition~\ref{prop:Chee} in Section~\ref{sec:rates}). 
\end{enumerate}

\section{Bounds on optimal code size}\label{sec:bounds}

We derive novel bounds for $C(m,L,d,w_s)$ and $S(m,L,d,w_s)$.
While bounds for the former were also discussed in \cite{Chee14_MCWC},
those results are insufficient to provide good bounds on the asymptotic rates $\gamma(L,\delta,w_s/L)$.
Among other bounds, we derive the Gilbert-Varshamov (GV) bound and the sphere-packing bound for both CSCCs and SECCs in this section, and
their respective asymptotic versions in Section \ref{sec:rates}.

\subsection{CSCC code size}
For an $(m,L,d,w_s)$-CSCC, it is easy to see from symmetry, via complementing bits in codewords, that we have the relation
\begin{equation}
C(m,L,d,w_s) = C(m,L,d,L-w_s) .
\end{equation}

Let $\mathcal{C}(m,L,w_s)$ denote the space of all binary words comprising of $m$ subblocks, each subblock having length $L$, with weight $w_s$ per subblock. For $\mathbf{x} \in \mathcal{C}(m,L,w_s)$, we define a ball centered at $\mathbf{x}$ and having radius $t$ as
\begin{equation}
\mathcal{B}_{\mathcal{C}}(\mathbf{x},t;m,L,w_s) \triangleq \{ \mathbf{y} \in \mathcal{C}(m,L,w_s) : d(\mathbf{x},\mathbf{y}) \le t \}
\end{equation}
The following lemma shows that the size of the CSCC ball, $|\mathcal{B}_{\mathcal{C}}(\mathbf{x},t;m,L,w_s)|$, is independent of choice of $\mathbf{x} \in \mathcal{C}(m,L,w_s)$. We will see later in Sec.~\ref{Sec:SECC_CodeSize} that this is not true for the space comprising of SECC words.
\begin{Lemma}
\label{Lemma:CSCC_Balls_Equal}
If $\mathbf{x}$ and $\tilde{\mathbf{x}}$ are two words in $\mathcal{C}(m,L,w_s)$, then $|\mathcal{B}_{\mathcal{C}}(\tilde{\mathbf{x}},t;m,L,w_s)| = |\mathcal{B}_{\mathcal{C}}(\mathbf{x},t;m,L,w_s)|$.
\end{Lemma}
\begin{IEEEproof}
For $1\le i \le m$, let $\mathbf{x}_{[i]}$ (resp. $\tilde{\mathbf{x}}_{[i]}$) denote the $i$th subblock of $\mathbf{x}$ (resp. $\tilde{\mathbf{x}}$). As  $\mathbf{x}_{[i]}$ and  $\tilde{\mathbf{x}}_{[i]}$ have constant weight $w_s$, there exists a permutation $\pi_i$ on $L$ letters such that $\tilde{\mathbf{x}}_{[i]} = \pi_i(\mathbf{x}_{[i]})$. Now, if $\pi$ denotes the permutation on $mL$ letter, induced by $\pi_i, 1\le i \le m$, defined as
\begin{equation}
\pi(\mathbf{x}) \triangleq [\pi_{1}(\mathbf{x}_{[1]}) \cdots \pi_m(\mathbf{x}_{[m]})] ,
\end{equation}
then, $\tilde{\mathbf{x}} = \pi(\mathbf{x})$. The proof is complete by observing that
\begin{equation}
\mathcal{B}_{\mathcal{C}}(\tilde{\mathbf{x}},t;m,L,w_s) = \left\{ \pi(\mathbf{y}) : \mathbf{y} \in \mathcal{B}_{\mathcal{C}}(\mathbf{x},t;m,L,w_s) \right\} .
\end{equation}
\end{IEEEproof}
In view of the above lemma, the size of CSCC ball is independent of the center word. Thus, we have the following GV bound for $C(m,L,d,w_s)$.
\begin{proposition}
	If $v \triangleq \min\{w_s, L-w_s\}$, then
\begin{equation}
C(m,L,d,w_s) \ge \frac{\binom{L}{w_s}^m}{\displaystyle \sum_{\substack{2(u_1+u_2+\cdots+u_m)\le d-1, \\ 0 \le u_i \le v}} ~\prod_{i=1}^m \binom{w}{u_i}\binom{L-w}{u_i}} \label{eq:CSCC_GV_v1}
\end{equation}
\label{prop:CSCC_GV}
\end{proposition}
\begin{IEEEproof}
Using standard Gilbert construction in the space $\mathcal{C}(m,L,w_s)$, we have the lower bound

\begin{equation}
C(m,L,d,w_s) \ge \frac{|\mathcal{C}(m,L,w_s)|}{|\mathcal{B}_{\mathcal{C}}(\mathbf{x},d-1;m,L,w_s)|} , \label{eq:CSCC_GV_v0}
\end{equation}
where $\mathbf{x}$ is any word in $\mathcal{C}(m,L,w_s)$, and $|\mathcal{C}(m,L,w_s)| = {\binom{L}{w_s}}^m$. From Lemma~\ref{Lemma:CSCC_Balls_Equal} we note that  $|\mathcal{B}_{\mathcal{C}}(\mathbf{x},d-1;m,L,w_s)|$ is independent of the choice of $\mathbf{x}$. The proposition then follows if we show that the denominator in~\eqref{eq:CSCC_GV_v1} is equal to $|\mathcal{B}_{\mathcal{C}}(\mathbf{x}, d-1; m, L,w)|$. Towards this, let $\mathbf{x}_{[i]}$ be the $i$th subblock of $\mathbf{x}$. Then the distance of $\mathbf{x}_{[i]}$ from any length $L$ binary vector of weight $w_s$ is always even, and the number of length $L$ binary vectors of weight $w_s$ at a distance $2 u_i$  from $\mathbf{x}_{[i]}$ is $\binom{w_s}{u_i} \binom{L-w_s}{u_i}$ when $0 \le u_i \le v$ (and 0 otherwise). Now, if $\mathbf{y} \in \mathcal{C}(m,L,w_s)$, and distance between $i$th subblocks of $\mathbf{x}$ and $\mathbf{y}$ is $2u_i$, then $\mathbf{y} \in \mathcal{B}_{\mathcal{C}}(\mathbf{x}, d-1; m, L,w)$ if and only if  $2 \sum_{i=1}^{m} u_i \le d-1$. Hence, the size of CSCC ball of radius $d-1$ is given by the denominator in~\eqref{eq:CSCC_GV_v1}. 
\end{IEEEproof}

The following proposition provides the sphere-packing bound for CSCCs.
\begin{proposition}
	If $v \triangleq \min\{w_s, L-w_s\}$ and $t \triangleq \lfloor (d-1)/2 \rfloor$, then
	\begin{equation}
	C(m,L,d,w_s) \le \frac{\binom{L}{w_s}^m}{\displaystyle \sum_{\substack{2(u_1+u_2+\cdots+u_m)\le t, \\ 0 \le u_i \le v}} ~\prod_{i=1}^m \binom{w}{u_i}\binom{L-w}{u_i}} \label{eq:CSCC_SP_v1}
	\end{equation}
	\label{prop:CSCC_SP}
\end{proposition}
\begin{IEEEproof}
The claim follows from the standard sphere-packing argument that for any $(m,L,d,w_s)$-CSCC, the balls of radius $t = \lfloor (d-1)/2 \rfloor$ around codewords should be non-intersecting, and the fact that the denominator in \eqref{eq:CSCC_SP_v1} is equal to $|\mathcal{B}_{\mathcal{C}}(\mathbf{x}, t; m, L,w)|$.
\end{IEEEproof}

\subsection{SECC code size} \label{Sec:SECC_CodeSize}
\subsubsection{Lower bounds on SECC code size}
Let $\mathcal{S}(m,L,w_s)$ denote the space of all binary words comprising of $m$ subblocks, each subblock having length $L$, with weight per subblock \emph{at least} $w_s$. For $\mathbf{x} \in \mathcal{S}(m,L,w_s)$, we define a ball centered at $x$ and having radius $t$ as
\begin{equation}
\mathcal{B}_{\mathcal{S}}(\mathbf{x},t;m,L,w_s) \triangleq \{ \mathbf{y} \in \mathcal{S}(m,L,w_s) : d(\mathbf{x},\mathbf{y}) \le t \}
\end{equation}
Unfortunately, in contrast to CSCCs, the size of $\mathcal{B}_{\mathcal{S}}(\mathbf{x},t;m,L,w_s)$ depends on $\mathbf{x}$. Take for example, $m=1$, $L=4$, $w_s=2$ and $t=1$.
We have that ${\cal B}_{\cal S}(0111, t; m, L,w_s)=\{0111,1111,0011,0101,0110\}$, while
${\cal B}_{\cal S}(1001, t; m, L,w)=\{1001,1101,1011\}$.

We denote the smallest and the average ball size in the SECC space as follows:
\begin{align}
|\mathcal{B}_{\mathcal{S}}^{\mathrm{min}}(t;m,L,w_s)| &\triangleq \min_{\mathbf{x} \in \mathcal{S}(m,L,w_s)} |\mathcal{B}_{\mathcal{S}}(\mathbf{x},t;m,L,w_s)| , \label{eq:MinSECCball} \\
|\mathcal{B}_{\mathcal{S}}^{\mathrm{avg}}(t;m,L,w_s)| &\triangleq \sum_{\mathbf{x} \in \mathcal{S}(m,L,w_s)} \frac{|\mathcal{B}_{\mathcal{S}}(\mathbf{x},t;m,L,w_s)|}{|\mathcal{S}(m,L,w_s)|} . \label{eq:AvgSECCball}
\end{align}

The total number of words in the SECC space $\mathcal{S}(m,L,w_s)$ are $\left(\sum_{i=w_s}^L \binom{L}{i}\right)^m$. Thus, the \emph{generalized Gilbert-Varshamov bound}~\cite[Thm.~4]{Gu93}, \cite{Tolhuizen97} (for spaces where balls with fixed radius and different centers may have different sizes) when applied to the SECC space $\mathcal{S}(m,L,w_s)$ gives us the following lower bound on $S(m,L,d,w_s)$.
\begin{proposition} \label{prop:SECC_GV_CodeSize}
	We have
	\begin{equation}
	S(m,L,d,w_s) \ge \frac{\left(\sum_{i=w_s}^L \binom{L}{i}\right)^m}{|\mathcal{B}_{\mathcal{S}}^{\mathrm{avg}}(d-1;m,L,w_s)|} . \label{eq:SECC_GV_v0}
	\end{equation}
\end{proposition}

The next proposition demonstrates how to construct SECCs from CSCCs.

\begin{proposition}\hfill

	\text{(i)} We have
	\begin{equation}
	S(m,L,d,w_s) \ge C(m,L,d,j), ~ \mathrm{for}~j\ge w_s.
	\end{equation}
	
	\text{(ii)} When $m \ge d$, we have
	\begin{equation}
	S(m,L,d,w_s) \ge \sum_{j=w_s}^L C(m,L,d,j) .
	\end{equation}

	(iii) When $m$ is an integer multiple of $d$, with $m = k d$,
	\begin{equation}
	S(m,L,d,w_s) \ge \left( \sum_{j=w_s}^L C(d,L,d,j) \right)^k.
	\end{equation}
\end{proposition}
\begin{IEEEproof}
	\text{(i)} A CSCC with fixed subblock weight $j \ge w_s$ is also an SECC having subblock weight at least $w_s$.
	
	\text{(ii)} If $\mathbf{s}$ and $\tilde{\mathbf{s}}$ are two CSCC sequences with $m$ subblocks, constant weight per subblock $j$ and $j+1$, respectively, then the Hamming distance between $\mathbf{s}$ and $\tilde{\mathbf{s}}$ is at least $m$.
	
	(iii) For $w_s \le j \le L$, let $\mathscr{C}_j$ denote an $(d,L,d,j)$-CSCC having size $C(d,L,d,j)$. Now construct a SECC code using $\mathscr{C}_j, ~w_s \le j \le L$, where each block comprising of $d$ consecutive subblocks is chosen from the set $\displaystyle \cup_{j=w_s}^L \mathscr{C}_j$. The resulting code has weight at least $w_s$ per subblock and minimum distance $d$. 
\end{IEEEproof}

The next proposition extends the concatenation approach~\cite{ForneyBook} for SECCs.

\begin{proposition}
	If $q \le H(L,d_1,w_s)$, then
	\begin{equation}
	S(m,L,d_1 d_2, w_s) \ge A_q(m,d_2).
	\end{equation}
\end{proposition}  
\begin{IEEEproof}
	Adapt the concatenated code construction scheme in~\cite[Prop.~4.1]{Chee14_MCWC} by replacing the constant weight inner code by a heavy weight inner code.
\end{IEEEproof}

We extend the Elias-Bassalygo bound (see for example, \cite[eq.~2.7]{McEliece77}) for SECCs. 

\begin{proposition}
	We have
	\begin{equation}
	S(m,L,d,w_s) \ge \frac{\left(\sum_{i=w_s}^L \binom{L}{i} \right)^m}{2^{mL}} A(mL,d) .
	\label{eq:SECC_EliasBound}
	\end{equation}
\end{proposition}
\begin{IEEEproof}
	Let $\mathscr{C}$ be a $(mL,d)$-code with $A(mL,d)$ codewords. Let $\mathbb{F}_2^{mL}$ denote the space of binary vectors of length $mL$, and $\mathbf{x} \in \mathbb{F}_2^{mL}$ be chosen so that $|\mathcal{S}(m,L,w_s) \cap (\mathbf{x} + \mathscr{C})|$ is maximal. Then
	\begin{align}
	S(m,L,d,w_s) &\ge |\mathcal{S}(m,L,w_s) \cap (\mathbf{x} + \mathscr{C})| \nonumber \\
	&\ge \frac{1}{2^{mL}} \sum_{\mathbf{y} \in \mathbb{F}_2^{mL}} |\mathcal{S}(m,L,w_s) \cap (\mathbf{y} + \mathscr{C})| \nonumber \\
	&= \frac{1}{2^{mL}} \sum_{\mathbf{y} \in \mathbb{F}_2^{mL}} \sum_{\mathbf{b} \in \mathcal{S}(m,L,w_s)} \sum_{\mathbf{c} \in \mathscr{C}} |\{\mathbf{b}\} \cap \{ \mathbf{y}+\mathbf{c}\}| \nonumber \\
	&= \frac{1}{2^{mL}} \sum_{\mathbf{b} \in \mathcal{S}(m,L,w_s)} \sum_{\mathbf{c} \in \mathscr{C}} 1 \nonumber \\
	&= \frac{|\mathcal{S}(m,L,w_s)| |\mathscr{C}|}{2^{mL}} . \nonumber 
	\end{align}
\end{IEEEproof}

\subsubsection{Upper bounds on SECC code size}

We next provide several upper bounds on the SECC code size, including the SECC sphere-packing bound (Prop.~\ref{prop:SECC_SP_CodeSize}) and the SECC Johnson type bound (Prop.~\ref{prop:SECC_Johnson_bound}). Observing that the balls of radius $t = \lfloor (d-1)/2 \rfloor$ around codewords should be non-intersecting in an $(m,L,d,w_s)$-SECC, we having the following sphere-packing upper bound on $S(m,L,d,w_s)$.
\begin{proposition} \label{prop:SECC_SP_CodeSize}	
	Let $t \triangleq \lfloor (d-1)/2 \rfloor$. Then, we have
	\begin{equation}
S(m,L,d,w_s) \le \frac{\left(\sum_{i=w_s}^L \binom{L}{i}\right)^m}{|\mathcal{B}_{\mathcal{S}}^{\mathrm{min}}(t;m,L,w_s)|} . \label{eq:SECC_SP_v0}
\end{equation}	
\end{proposition}

As discussed earlier, for a given radius $t$, different SECC balls may have different sizes, depending on the center word. In view of this, note that the SECC sphere-packing upper bound~\eqref{eq:SECC_SP_v0} is obtained by considering the smallest ball size of radius $t$. The \emph{generalized sphere-packing bound}, for spaces where different balls of same radius have different sizes, was investigated in \cite{Fazeli15,Cullina16}. 
However, it is unclear if the techniques in \cite{Fazeli15,Cullina16} are able to yield tighter \emph{asymptotic} upper bound than that given in the next section via Theorem~\ref{thm:SECC_SP_rate}. 

Furthermore, we point out that the average sphere-packing value is \emph{not} an upper bound for 
the code size of SECCs.
Specifically, for a $t$-error-correcting code, the \emph{average sphere-packing value} was defined in \cite{Fazeli15} to be the ratio of size of the space, to the average ball size of radius $t$. It was observed that for many spaces, this average sphere-packing value is an upper bound for the optimal code size.

However, we now show that there exist SECC spaces where the average sphere-packing value is \emph{not} an upper bound on the optimal code size. Towards this, consider the SECC space, $\mathcal{S}(m,L,w_s)$, corresponding to $m=1$, $L=3$, and $w_s=1$. Here, the size of space, $|\mathcal{S}(m,L,w_s)|$, is 7 while the average ball size, $|\mathcal{B}_{\mathcal{S}}^{\mathrm{avg}}(t;m,L,w_s)|$, corresponding to $t=1$ is equal to 25/7. In this case, the \emph{average sphere-packing value}, for a single error correcting code, is 49/25. But this value is readily seen to be strictly less than the size of the SECC code $\mathscr{C} = \{100, 011\}$.

Other upper bounds on the optimal SECC code size are discussed next. The inequality
\begin{equation}
S(m,L,d,w_1) \le S(m,L,d,w_2), \hfill \mathrm{if~} w_1 > w_2 ,
\end{equation}
is immediate from the definition of SECC. The following proposition employs puncturing to bound $S(m,L,d,w_s)$.
\begin{proposition}
	If $d>m$, then
	\begin{equation}
	S(m,L,d,w_s) \le S(m,L-1,d-m,w_s-1).
	\end{equation}
\end{proposition}
\begin{IEEEproof}
	Consider an $(m,L,d,w_s)$-SECC with $S(m,L,d,w)$ codewords. Arrange each codeword in a $m\times L$ matrix. Since $d>m$, puncturing any fixed column in each codeword results in a $(m,L-1,d-m,w_s-1)$-SECC with $S(m,L,d,w)$ codewords.
\end{IEEEproof}

We now present a Johnson type bound~\cite{Johnson62,Johnson72} for SECCs, which provides an upper bound on $S(m,L,d,w)$.
Towards this, we consider a generalization of SECC where different subblocks in a codeword may have different length and weight constraints. Let $T(m, [L_1, \ldots, L_m], d, [w_1, \ldots, w_m])$ denote the largest size of a binary code where each codeword has $m$ subblocks, the $i$th subblock has length $L_i$ and weight at least $w_i$, and the minimum distance of the code is $d$. Here, the length of each codeword is $n = \sum_{i=1}^m L_i$. 

Now, let $\mathscr{C}$ be such a generalized code of size $T(m, [L_1, \ldots, L_m], d, [w_1, \ldots, w_m])$. Consider a matrix with $n$ columns, whose rows comprise of the $T(m, [L_1, \ldots, L_m], d, [w_1, \ldots, w_m])$ codewords of $\mathscr{C}$. By focusing on the $i$th subblock of each codeword, we observe that there exists a column having at least $T(m, [L_1, \ldots, L_m], d, [w_1, \ldots, w_m]) \cdot (w_i/L_i)$ ones, and denote such a column as $l$. Pick a subcode of $\mathscr{C}$ where each codeword has a 1 in the $l$-th position. Delete the $l$-th component in the subcode to obtain

\begin{equation}
T(m, [L_1, \ldots, L_m], d, [w_1, \ldots, w_m]) \le \frac{L_i}{w_i} T(m, [L_1, \ldots, L_i - 1, \ldots L_m], d, [w_1, \ldots, w_i - 1, \ldots, w_m]) .
\label{eq:GeneralizedSECC_UpperBound}
\end{equation}

By varying $i$ from $1$ to $m$ and recursively applying \eqref{eq:GeneralizedSECC_UpperBound},
\begin{equation}
T(m, [L_1, \ldots, L_m], d, [w_1, \ldots, w_m]) \le \left( \prod_{i=1}^{m} \frac{L_i}{w_i} \right) T(m, [L_1 - 1, \ldots, L_m - 1], d, [w_1 - 1, \ldots, w_m - 1]) .
\label{eq:GeneralizedSECC_UpperBound_v2}
\end{equation}

Specializing \eqref{eq:GeneralizedSECC_UpperBound_v2} to the case when each $L_i = L$ and $w_i = w_s$, we obtain the following Johnson type bound for SECCs.

\begin{proposition} \label{prop:SECC_Johnson_bound}
	We have
	\begin{equation}
	S(m,L,d,w_s) \le \frac{L^m}{w_s^m} S(m,L-1,d,w_s-1) .
	\label{eq:SECC_UpperBound}
	\end{equation}
\end{proposition}

We next present bounds on the asymptotic rate for CSCCs and SECCs.

\section{Asymptotic Bounds on Rate}\label{sec:rates}
The asymptotic rate for subblock constrained codes may be studied in scenarios where the number of subblocks $m$, or the subblock length $L$, or both, tend to infinity. The following proposition states that the asymptotic rate of CSCC is equal to that of CWC when the subblock length $L$ tends to infinity, which is not surprising as the subblock constraint fades asymptotically.


\begin{proposition}\label{prop:Chee}
	For any positive integer $m$ and $0\le \delta, \omega\le 1$, 
	\begin{equation}\label{eq:fixed-m-cscc}
	\lim_{L\to\infty} \frac{\log C(m,L, \floor{\delta mL},\floor{\omega L})}{mL}=\alpha(\delta,\omega).
	\end{equation}
\end{proposition}
\begin{IEEEproof}
	We have the inequality
	\begin{equation}
	\lim_{L\to\infty} \frac{\log C(m,L, \floor{\delta mL},\floor{\omega L})}{mL} \le \alpha(\delta,\omega) , \label{eq:CSCC_Linf_rate_UB}
	\end{equation}
	as $C(m,L,\floor{\delta mL},\floor{\omega L}) \le A(mL,\floor{\delta mL},m\floor{\omega L})$.
	On the other hand, from \cite[Lemma~6.1]{Chee14_MCWC}, we have that
	\begin{equation}
	C(m,L,\floor{\delta mL},\floor{\omega L}) \ge \frac{{\binom{L}{\omega L}}^m}{\binom{m L}{\omega m L}} A(mL,\floor{\delta mL},m\floor{\omega L}). \label{eq:CSCC_Linf_LB}
	\end{equation}
	If $h(\cdot)$ denotes the binary entropy function, then
	\begin{equation}
	\lim_{L \to \infty} \frac{1}{mL} \log \frac{{\binom{L}{\omega L}}^m}{\binom{m L}{\omega m L}} = h(\omega) - h(\omega) = 0,
	\end{equation}
	and hence using \eqref{eq:CSCC_Linf_LB} we have
	\begin{equation}
	\lim_{L\to\infty} \frac{\log C(m,L, \floor{\delta mL},\floor{\omega L})}{mL} \ge \alpha(\delta,\omega) . \label{eq:CSCC_Linf_rate_LB}
	\end{equation}
	The proof is complete by combining \eqref{eq:CSCC_Linf_rate_UB} and \eqref{eq:CSCC_Linf_rate_LB}.
\end{IEEEproof}
	
Note that \eqref{eq:fixed-m-cscc} also holds when $m \to \infty$.  Asymptotic rate results were also presented in \cite{Chee14_MCWC}.   However, there were some inconsistencies in the definition of the asymptotic CSCC rate and the resulting claim in \cite[Prop.~6.1]{Chee14_MCWC} was incorrect.  Proposition \ref{prop:Chee} above provides a correction. The inconsistency in the rate definition in \cite{Chee14_MCWC} also renders \cite[Thm.~6.3]{Chee14_MCWC} incorrect, whose proof also contained some anomalies.

By combining Thm.~\ref{thm:Bachoc} and Prop.~\ref{prop:Chee}, we obtain the following proposition on the SECC asymptotic rate in the scenario where the subblock length $L$ tends to infinity.
\begin{proposition}\label{prop:BachocChee}
	For any positive integer $m$, and $0\le \delta\le 1$, $1/2\le \omega\le 1$, we have 
	\begin{equation}\label{eq:fixed-m-secc}
	\lim_{L\to\infty} \frac{\log S(m,L, \floor{\delta mL},\floor{\omega L})}{mL}=\eta(\delta,\omega).
	\end{equation}
\end{proposition}

In the remainder of the paper, we fix the relative distance $\delta$, the subblock length $L$, 
and the parameter $w_s$, and provide estimates of the asymptotic rates for CSCCs and SECCs 
as the number of blocks $m$ tend to infinity. The motivation for fixing $L$ to relatively small values comes from the application of CSCCs and SECCs to \emph{simultaneous energy and information transfer}~\cite{Tandon16_CSCC_TIT}. Here, it can be shown that if the weight of each subblock is sufficiently high, then a receiver with limited energy storage will not suffer from energy outage when the subblock length is less than a certain threshold~\cite{Tandon16_CSCC_TIT,Tandon16_ISIT}.

\subsection{CSCC Rate}

Recall the definitions of $\gamma(L,\delta,w_s/L)$ and $\alpha(\delta, w_s/L )$
given by \eqref{eq:CSCC_RateDef} and \eqref{eq:CWC_RateDef}. 
Furthermore, these quantities are related via the following inequality
\begin{equation}
\gamma(L,\delta,w_s/L) \le \alpha(\delta,w_s/L) . \label{eq:CSCC_CWC_rate}
\end{equation}
The following proposition shows that for the case when $L=2$ and $w_s=1$, the CSCC rate $\gamma(L,\delta,w_s/L)$ is \emph{strictly} less than $\alpha(\delta,w_s/L)$ when $0 < \delta < 1/2$.
\begin{proposition}
	We have 
	\begin{equation}
	\gamma(2,\delta,1/2) = \frac{1}{2} \alpha(\delta,1/2) . \label{eq:CSCC_Rate_L2}
	\end{equation}
\end{proposition}
\begin{IEEEproof}
	It was shown in \cite[Cor.~4.2]{Chee14_MCWC} that \[C(m,2,2d,1)=A(m,d).\] 
	Then \eqref{eq:CSCC_Rate_L2} follows immediately from the definitions of asymptotic rates.
\end{IEEEproof}

Since $\alpha(\delta,1/2) = \alpha(\delta)$ \cite{McEliece77}, the relation in \eqref{eq:CSCC_Rate_L2} can alternately be expressed as $\gamma(2,\delta,1/2) = (1/2) \alpha(\delta)$. Now, from the GV bound for general binary codes~\cite{McEliece77}, we know that $\alpha(\delta) > 0$ for $0 < \delta < 0.5$, while from the asymptotic Plotkin bound~\cite{Plotkin60} for binary codes, we have $\alpha(\delta) = 0$ for $\delta \ge 0.5$. Thus, from \eqref{eq:CSCC_Rate_L2}, it follows that the inequality in~\eqref{eq:CSCC_CWC_rate} is strict for the case when $L=2$, $w_s=1$, and $0 < \delta < 0.5$. 

In general, for $L\ge 3$, define
\begin{equation}
\delta^* \triangleq 2 \left(\frac{w_s}{L}\right)\left(1 - \frac{w_s}{L}\right). \label{eq:delta_def}
\end{equation}
From the MRRW bound for constant weight codes~\cite[Eq.~(2.16)]{McEliece77}, we have
\begin{equation}
\alpha(\delta,w_s/L) = 0, ~\mathrm{if}~\delta \ge \delta^* . \label{eq:CWC_Rate0}
\end{equation}
From \eqref{eq:CSCC_CWC_rate} and \eqref{eq:CWC_Rate0}, it follows that 
\begin{equation}
\gamma(L,\delta,w_s/L) = 0, ~\mathrm{if}~ \delta \ge \delta^* . \label{eq:CSCC_Rate0}
\end{equation}
Therefore, we are interested in determining $\gamma(L,\delta,w_s/L)$ for $\delta< \delta^*$.
In fact, we will show that the inequality~\eqref{eq:CSCC_CWC_rate} is strict for relatively small values of $L$ and $\delta$. To this end, Theorem~\ref{th:CSCC_Rate_GV} presents a lower bound for $\gamma(L,\delta,w_s/L)$ using the GV bound for $C(m,L,d,w_s)$ when $\delta < \delta^*$. 
The following lemmas will be used towards proving this theorem.

\begin{Lemma} \label{Lemma:BinomLogConcave}
	For fixed positive integers $m$, $n$ and $z$, let $k_i$, with $1 \le i \le m$, be integers which satisfy $0 \le k_i \le n$, $\sum_{i=1}^m k_i = z$. Then we have the inequality
	\begin{equation}
	\prod_{i=1}^{m} \binom{n}{k_i} \le {\binom{n}{\lfloor z/m \rfloor}}^{m_1}  {\binom{n}{\lceil z/m \rceil}}^{m - m_1},\label{eq:BinomLogConcave}
	\end{equation}
	where $m_1 = m \lceil z/m \rceil - z$.
\end{Lemma}
\begin{IEEEproof}
	Follows from log-concavity of the binomial coefficients~\cite{Stanley89}.
\end{IEEEproof}

\begin{Lemma} \label{Lemma:BinomProdIneq}
	For $0 < k \le w_s(L-w_s)/L$, we have the inequality
	\begin{equation}
	\binom{w_s}{k} \binom{L-w_s}{k} > \binom{w_s}{k-1} \binom{L-w_s}{k-1} . \label{eq:BinomProdIneq}
	\end{equation} 
\end{Lemma}
\begin{IEEEproof}
We have
\begin{align*}
\frac{\binom{w_s}{k} \binom{L-w_s}{k}}{\binom{w_s}{k-1} \binom{L-w_s}{k-1}} &= \frac{\left(w_s - (k-1)\right)\left((L-w_s) - (k-1)\right)}{k^2} \\
&\overset{(a)}{\ge} \frac{L k - L(k-1) + (k-1)^2}{k^2} \\
&= \frac{(L-2k) + (k^2+1)}{k^2} \\
&\overset{(b)}{>} 1 ,
\end{align*}	
where $(a)$ follows because $w_s(L-w_s) \ge L k$, and 
$(b)$ follows from the fact that $k \le \min\{w_s,L-w_s\}\le L/2$. 
\end{IEEEproof}


\begin{theorem}[Asymptotic GV bound for CSCCs] \label{th:CSCC_Rate_GV}
For $0 < \delta < \delta^*$, we have
\begin{equation}
\gamma(L,\delta,w_s/L) \ge \gamma_{GV}(L,\delta,w_s/L) ,
\end{equation}
where $\gamma_{GV}(L,\delta,w_s/L)$ is defined as follows
\begin{enumerate}[a)]
	\item For $L=2$,
	\begin{equation} 
	\gamma_{GV}(2,\delta,1/2) \triangleq \frac{1}{2} (1 - h(\delta)) . \label{eq:CSCC_GV_L2}
	\end{equation}
	\item For $L>2$,
	\begin{align}
	\gamma_{GV}(L,\delta,w_s/L) &\triangleq \frac{1}{L}\log \binom{L}{w_s} -  
	\left(\frac{\lceil u \rceil - u}{L}\right) \log \binom{w_s}{\lfloor u \rfloor} - \left(\frac{1+u- \lceil u \rceil}{L}\right) \log \binom{w_s}{\lceil u \rceil} \nonumber \\
	&- \left(\frac{\lceil u \rceil - u}{L}\right) \log \binom{L - w_s}{\lfloor u \rfloor} - \left(\frac{1 + u - \lceil u \rceil}{L}\right) \log \binom{L - w_s}{\lceil u \rceil} \nonumber \\
	&- \min\{\theta(L,w_s), \,\phi(L,\delta)\} ,
	\label{eq:CSCC_GV}
	\end{align}
	where $u \triangleq \delta L/2$, and
	\begin{align*}
	\theta(L,w_s) &\triangleq \frac{1}{L} \log \left(\min\{w_s, L-w_s\} + 1\right) \\
	\phi(L,\delta) &\triangleq \left(\frac{1}{L} + \frac{\delta}{2} \right) h\left(\frac{1}{1+\delta L/2}\right) .
	\end{align*}
\end{enumerate}
\end{theorem}
\begin{IEEEproof}
The claim for $L=2$ follows from~\eqref{eq:CSCC_Rate_L2} and the GV bound for general binary codes.

For establishing the result for $L>2$, we use Proposition~\ref{prop:CSCC_GV}. The challenge here is to provide an appropriate upper bound on the CSCC ball size of radius $d-1$, $|\mathcal{B}_{\mathcal{C}}(\mathbf{x},d-1;m,L,w_s)|$ (the denominator in~\eqref{eq:CSCC_GV_v1}). From~\eqref{eq:delta_def} it follows that conditions $\delta < \delta^*$ and $d=\lfloor m L \delta \rfloor$ imply
\begin{equation}
d-1 \, < \, 2 m\,w_s(L-w_s)/L . \label{eq:Bound_d}
\end{equation}
Further, if $v \triangleq \min\{w_s,L-w_s\}$ and $t \triangleq \lfloor (d-1)/2 \rfloor$, we note that $|\mathcal{B}_{\mathcal{C}}(\mathbf{x},d-1;m,L,w_s)|$ is equal to
\begin{align}
&1 + \sum_{\tau = 1}^{t} ~\sum_{\substack{u_1+u_2+\cdots+u_m = \tau, \\ 0 \le u_i \le v}} \left[\prod_{i=1}^m \binom{w_s}{u_i}\right] \left[\prod_{i=1}^m \binom{L-w_s}{u_i}\right] \nonumber \\
&\overset{\text{(i)}}{\le} 1 + \sum_{\tau = 1}^{t} ~\sum_{\substack{u_1+u_2+\cdots+u_m = \tau, \\ 0 \le u_i \le v}} Q_{\tau} , \label{{eq:UpperBoundCSCCball}}
\end{align} 	
where $\text{(i)}$ follows from Lemma~\ref{Lemma:BinomLogConcave} and the definition
\begin{align}
Q_{\tau} &\triangleq \left[\binom{w_s}{\lfloor \tau/m \rfloor}\binom{L - w_s}{\lfloor \tau/m \rfloor}\right]^{m_\tau} \left[\binom{w_s}{\lceil \tau/m \rceil}\binom{L - w_s}{\lceil \tau/m \rceil}\right]^{m - m_\tau} \label{eq:Q_def} \\
m_\tau &\triangleq m \lceil \tau/m \rceil - \tau .
\end{align}
From~\eqref{eq:Bound_d} and relations $1 \le \tau \le t$, with $t = \lfloor (d-1)/2 \rfloor$, it follows that $\tau/m < w_s(L-w_s)/L$. Now, applying Lemma~\ref{Lemma:BinomProdIneq} we note that $Q_\tau$ is a non-decreasing function of $\tau$. Hence, using~\eqref{{eq:UpperBoundCSCCball}} we get
\begin{equation}
|\mathcal{B}_{\mathcal{C}}(\mathbf{x},d-1;m,L,w_s)| < Q_t \sum_{\substack{u_1+u_2+\cdots+u_m \le t, \\ 0 \le u_i \le v}} 1 , \label{eq:UpperBoundCSCCball_v1}
\end{equation}
where the summation term simply denotes the number of tuples $(u_1,u_2,\ldots,u_m)$ under two constraints: $u_1+\cdots+u_m \le t$ and $0 \le u_i \le v$. This summation term can hence be upper bounded as
\begin{equation}
\sum_{\substack{u_1+u_2+\cdots+u_m \le t, \\ 0 \le u_i \le v}} 1 ~\le~ \min\left\{(v+1)^m , ~\binom{t+m}{m} \right\}  , \label{eq:CSCC_partitions}
\end{equation}
where $(v+1)^m$ is obtained by counting the tuples $(u_1,u_2,\ldots,u_m)$ which only satisfy the constraint $0 \le u_i \le v$, while $\binom{t+m}{m}$ is obtained using the concept of \emph{weak compositions}~\cite[Chap.~2]{MiklosBonaBook} by counting the tuples which satisfy the constraints $u_1+\cdots+u_m \le t$ and $u_i \ge 0$ (with no upper bound on $u_i$ for $1\le i \le m$). Combining~\eqref{eq:UpperBoundCSCCball_v1} and \eqref{eq:CSCC_partitions},
\begin{equation}
|\mathcal{B}_{\mathcal{C}}(\mathbf{x},d-1;m,L,w_s)| < Q_t \, \min\left\{(v+1)^m , ~\binom{t+m}{m} \right\} , \label{eq:BoundCSCCball_v2}
\end{equation}
where $Q_t$ is obtained by substituting $\tau = t$ in~\eqref{eq:Q_def}. Finally, using~\eqref{eq:CSCC_GV_v0}, \eqref{eq:CSCC_RateDef}, and \eqref{eq:BoundCSCCball_v2}, we observe that $\gamma(L,\delta,w_s/L)$ is lower bounded by
\begin{equation}
 \frac{1}{L}\log \binom{L}{w_s} - \lim_{m \to \infty} \frac{1}{mL}\log Q_t - \min\left\{ \frac{1}{L} \log (v+1) , \lim_{m \to \infty} \frac{1}{mL} \log \binom{t+m}{m} \right\} . \label{eq:LowerBoundGamma}
\end{equation}
It can be verified that
\begin{equation}
\lim_{m \to \infty} \frac{1}{mL} \log \binom{t+m}{m} = \left(\frac{1}{L} + \frac{\delta}{2} \right) h\left(\frac{1}{1+\delta L/2}\right) . \label{eq:PhiTerm}
\end{equation}
Now, the $t/m$ term in the expression for $Q_t$~\eqref{eq:Q_def} is equal to $\delta L/2 =: u$, and it follows using~\eqref{eq:PhiTerm} that the lower bound on $\gamma(L,\delta,w_s/L)$ given by~\eqref{eq:LowerBoundGamma} simplifies to the expression on the right hand side in~\eqref{eq:CSCC_GV}.
\end{IEEEproof}	
\emph{Remark}: In general, we have $\phi(L,\delta) < \theta(L,w_s)$ when $w_s/L$ does not deviate significantly from $1/2$. On the other hand, $\theta(L,w_s)$ may be less than $\phi(L,\delta)$ when $w_s$ is extremely close to $L$ (or 0). In particular, $\theta(L,L-1) < \phi(L,\delta)$ if $\delta \ge \frac{2}{3L}$.

The following theorem presents the asymptotic sphere-packing upper bound on $\gamma(L,\delta,w_s/L)$ when $\delta < \delta^*$.
\begin{theorem}[Asymptotic sphere-packing bound for CSCCs]
For $0 < \delta < \delta^*$, we have
\begin{equation}
\gamma(L,\delta,w_s/L) \le \gamma_{SP}(L,\delta,w_s/L) ,
\end{equation}
where $\gamma_{SP}(L,\delta,w_s/L)$ is defined as
\begin{align}
\gamma_{SP}(L,\delta,w_s/L) &\triangleq \frac{1}{L}\log \binom{L}{w_s} -  
\left(\frac{\lceil\tilde{u}\rceil - \tilde{u}}{L}\right) \log \binom{w_s}{\lfloor\tilde{u}\rfloor} - \left(\frac{1+\tilde{u}- \lceil\tilde{u}\rceil}{L}\right) \log \binom{w_s}{\lceil\tilde{u}\rceil} \nonumber \\
&- \left(\frac{\lceil\tilde{u}\rceil - \tilde{u}}{L}\right) \log \binom{L - w_s}{\lfloor\tilde{u}\rfloor} - \left(\frac{1 + \tilde{u} - \lceil\tilde{u}\rceil}{L}\right) \log \binom{L - w_s}{\lceil\tilde{u}\rceil} \nonumber \\
&- \frac{1}{L} h(\lceil\tilde{u}\rceil - \tilde{u}) ,
\label{eq:CSCC_SP}
\end{align}
where $\tilde{u} \triangleq \delta L/4$.
\end{theorem}
\begin{IEEEproof}
For proving the claim, we apply Proposition~\ref{prop:CSCC_SP} while providing an appropriate lower bound on the CSCC ball size of radius $t$, $|\mathcal{B}_{\mathcal{C}}(\mathbf{x},t;m,L,w_s)|$ (the denominator in~\eqref{eq:CSCC_SP_v1}), where $t = \lfloor (d-1)/2 \rfloor$ and distance $d$ scales as $d = \lfloor m L \delta \rfloor$. If we define $v \triangleq \min\{w_s, L-w_s\}$, $\tilde{t} \triangleq \lfloor t/2 \rfloor$, and $\tilde{v} \triangleq \tilde{t}/m$, then we note that $\tilde{v}$ satisfies the following inequality
\begin{align}
0 < \tilde{v} < \frac{d}{4m} \le \frac{L \delta}{4} \overset{\text{(i)}}{<} \frac{v}{2} , \label{eq:tilde_u_inequality}
\end{align}
where $\text{(i)}$ follows from the inequality $\delta < \delta^*$. Further, we have
	\begin{align}
	|\mathcal{B}_{\mathcal{C}}(\mathbf{x},t;m,L,w_s)| &= 1 + \sum_{\tau = 1}^{\tilde{t}} ~\sum_{\substack{u_1+u_2+\cdots+u_m = \tau, \\ 0 \le u_i \le v}} ~\prod_{i=1}^m \binom{w_s}{u_i} \binom{L-w_s}{u_i} \nonumber \\
	&\overset{\text{(ii)}}{>} \sum_{\substack{u_1+u_2+\cdots+u_m = \tilde{t}, \\ \lfloor \tilde{v} \rfloor \le u_i \le \lceil \tilde{v} \rceil}} ~\prod_{i=1}^m \binom{w_s}{u_i} \binom{L-w_s}{u_i} \nonumber \\
	&=  \binom{m}{\tilde{m}} \left[\binom{w_s}{\lfloor \tilde{v} \rfloor}\binom{L - w_s}{\lfloor \tilde{v} \rfloor}\right]^{\tilde{m}} \left[\binom{w_s}{\lceil \tilde{v} \rceil}\binom{L - w_s}{\lceil \tilde{v} \rceil}\right]^{m - \tilde{m}}, \label{eq:LowerBoundCSCCball}
	\end{align} 	
	where $\text{(ii)}$ follows using~\eqref{eq:tilde_u_inequality} (as the constraint $\lfloor \tilde{v} \rfloor \le u_i \le \lceil \tilde{v} \rceil$ is stricter than the constraint $0 \le u_i \le v$), and $\tilde{m}$ in~\eqref{eq:LowerBoundCSCCball} is defined as $\tilde{m} \triangleq m \lceil \tilde{t}/m \rceil - \tilde{t}$. Note that asymptotically we get the following limits
	\begin{align}
	\lim_{m \to \infty} \tilde{v} &= \lim_{m \to \infty} \frac{1}{m} \left\lfloor \frac{1}{2} \left\lfloor \frac{d-1}{2} \right\rfloor \right\rfloor = \frac{\delta L}{4} = \tilde{u} , \label{eq:CSCC_limit_tilde_v} \\
	\lim_{m \to \infty} \frac{\tilde{m}}{m} &= \lim_{m \to \infty} \left(\lceil \tilde{v} \rceil - \tilde{v}\right) = \lceil \tilde{u} \rceil - \tilde{u} . \label{eq:CSCC_limit_tilde_m}
	\end{align}
	Using \eqref{eq:LowerBoundCSCCball}, \eqref{eq:CSCC_limit_tilde_v}, and \eqref{eq:CSCC_limit_tilde_m}, we have 
	\begin{align}
	 \lim_{m \to \infty} \frac{1}{mL} \log |\mathcal{B}_{\mathcal{C}}(\mathbf{x},t;m,L,w_s)| &> \frac{1}{L} h\left(\lceil \tilde{u} \rceil - \tilde{u}\right)  + \left(\frac{\lceil\tilde{u}\rceil - \tilde{u}}{L}\right) \log \binom{w_s}{\lfloor\tilde{u}\rfloor} \nonumber \\ 
	 &+ \left(\frac{1+\tilde{u}- \lceil\tilde{u}\rceil}{L}\right) \log \binom{w_s}{\lceil\tilde{u}\rceil} 
	+  \left(\frac{\lceil\tilde{u}\rceil - \tilde{u}}{L}\right) \log \binom{L - w_s}{\lfloor\tilde{u}\rfloor} \nonumber \\
	&+ \left(\frac{1 + \tilde{u} - \lceil\tilde{u}\rceil}{L}\right) \log \binom{L - w_s}{\lceil\tilde{u}\rceil} . \label{eq:CSCC_BoundLogBallSize}
	\end{align}
	The theorem is proved by combining Proposition~\ref{prop:CSCC_SP},~\eqref{eq:CSCC_RateDef},  and~\eqref{eq:CSCC_BoundLogBallSize}. 
\end{IEEEproof}
\emph{Remark}: For $L=2, w_s=1$, we have $\gamma_{SP}(2,\delta,0.5) = 0.5(1-h(\delta/2))$, which also follows from $\gamma(2,\delta,1/2) = (1/2) \alpha(\delta)$ and then applying the standard sphere-packing bound for unconstrained binary codes.

\emph{Remark}: For $0 < \delta < \min\left\{ \delta^* , \frac{4}{L}\right\}$, \eqref{eq:CSCC_SP} simplifies to
\begin{equation}
\gamma_{SP}(L,\delta,w_s/L) = \frac{1}{L}\log \binom{L}{w_s} - \frac{\delta}{4} \log \left(w_s (L-w_s)\right) - \frac{1}{L} h\left(\frac{\delta L}{4}\right) \label{eq:CSCC_SP_v2}
\end{equation}

\subsection{SECC Rate}

Recall the definitions of $\sigma(L,\delta,w_s/L)$ and $\eta(\delta, w_s/L )$
given by \eqref{eq:SECC_RateDef} and \eqref{eq:HWC_RateDef}. 
We have the following inequality  

\begin{equation}
\sigma(L,\delta, w_s/L) \le \eta(\delta, w_s/L) . \label{eq:SECC_HWC_Rate}
\end{equation}
The gap $\eta(\delta, w_s/L) - \sigma(L,\delta, w_s/L)$ denotes the rate penalty on HWC due to the additional constraint on sufficient weight within every \emph{subblock duration}. 
The asymptotic rates of HWCs were studied in~\cite{Bachoc11} where Theorem \ref{thm:Bachoc} was established. Therefore, it follows that for $w_s \ge L/2$ we have
\begin{equation}
\sigma(L,\delta, w_s/L) = 0, ~\mathrm{if}~ \delta \ge \delta^* .
\end{equation}
In the following, we present the asymptotic GV bound and the sphere-packing bound on $\sigma(L,\delta, w_s/L)$.

\begin{proposition}[Asymptotic GV bound for SECCs] \label{prop:SECC_GV_rate}
We have $\sigma(L,\delta,w_s/L) \ge \sigma_{GV}(L,\delta,w_s/L)$ where
\begin{equation}
\sigma_{GV}(L,\delta,w_s/L) \triangleq \frac{1}{L}\log \left( \sum_{j=w_s}^L \binom{L}{j} \right) - h(\delta) . \label{eq:SECC_GV_rate}
\end{equation}
\end{proposition}
\begin{IEEEproof}
A simple upper bound on the average SECC ball size of radius $d-1$ is given by
\begin{equation}
|\mathcal{B}_{\mathcal{S}}^{\mathrm{avg}}(d-1;m,L,w_s)| \le \sum_{i=1}^{d-1} \binom{m L}{i} . \label{eq:UpperBoundSECCball}
\end{equation}
Using Proposition~\ref{prop:SECC_GV_CodeSize} and \eqref{eq:UpperBoundSECCball}, we get
\begin{equation}
S(m,L,d,w_s) \ge \frac{\left(\sum_{j=w_s}^L \binom{L}{j}\right)^m}{\sum_{i=1}^{d-1} \binom{m L}{i}} . \label{eq:SECC_GV_v1}
\end{equation}
The proposition now follows by combining~\eqref{eq:SECC_RateDef} and \eqref{eq:SECC_GV_v1}. 
\end{IEEEproof}
The above proposition presents a lower bound on $\sigma(L,\delta, w_s/L)$. Next, in Theorem~\ref{thm:SECC_SP_rate} we present the sphere-packing upper bound on $\sigma(L,\delta, w_s/L)$ for relatively small values of $\delta$. We will use the following lemma towards proving this theorem.

\begin{Lemma} \label{Lemma:SECC_subblocks_d12}
Let $\mathbf{x}_i$ be a binary vector of length $L$ whose weight $\tilde{w_s}$ satisfies $\tilde{w_s} \ge w_s$. Then the number of binary vectors with length $L$, weight at least $w_s$, which are at a distance of either 1 or 2 from $\mathbf{x}_i$ is lower bounded by $(L-w_s)(w_s+1)$.
\end{Lemma}

\begin{IEEEproof}
Let $N_1$ (resp. $N_2$) be the number of $L$ length vectors of weight at least $w_s$ which are at a distance 1 (resp. 2) from $\mathbf{x}_i$. We consider three different cases:
\begin{enumerate}
	\item $\tilde{w_s} = w_s$: ~In this case $N_1 = L - w_s$. If $(L-w_s) \ge 2$, then $N_2 = (L-w_s)w_s + \binom{L-w_s}{2}$, else $N_2 = (L-w_s)w_s$.
	
	\item $\tilde{w_s} = w_s + 1$: ~In this case $N_1 = L$. If $(L-w_s) \ge 2$, then $N_2 = (L-w_s)w_s + \binom{L-w_s}{2}$, else $N_2 = (L-w_s)w_s$.
	
	\item $\tilde{w_s} \ge w_s +2$: ~In this scenario, $N_1 = L$ and $N_2 = \binom{L}{2}$.
\end{enumerate}
For all the above three cases, it can easily be verified that $N_1 + N_2 \ge (L-w_s)(w_s+1)$.
\end{IEEEproof}
 
\begin{theorem}[Asymptotic sphere-packing bound for SECCs] \label{thm:SECC_SP_rate}
For $0 < \delta < \min\{\delta^*,4/L\}$, we have $\sigma(L,\delta,w_s/L) \le \sigma_{SP}(L,\delta,w_s/L)$ where 
\begin{equation}
\sigma_{SP}(L,\delta,w_s/L) \triangleq \frac{1}{L} \log \left( \sum_{j=w_s}^L \binom{L}{j} \right) - \frac{1}{L} h\left(\frac{L \delta}{4}\right) - \frac{\delta}{4} \log \left((L-w_s)(w_s+1)\right) . \label{eq:SECC_SP_rate}
\end{equation}
\end{theorem}
\begin{IEEEproof}
The theorem will be proved by using Prop.~\ref{prop:SECC_SP_CodeSize} and providing a lower bound on $|\mathcal{B}_{\mathcal{S}}^{\mathrm{min}}(t;m,L,w_s)|$ where $t = \lfloor (d-1)/2 \rfloor$ and distance $d$ scales as $\lfloor m L \delta \rfloor$. We define $\tilde{m} \triangleq \lfloor t/2 \rfloor$ and note that the constraint $\delta < 4/L$ implies that $\tilde{m} < m$. For a given $\mathbf{x} \in \mathcal{S}(m,L,w_s)$, let $\mathbf{x}_{[j]}$ denote the $j$-th subblock of $\mathbf{x}$, i.e. $\mathbf{x} = (\mathbf{x}_{[1]} \, \mathbf{x}_{[2]} \cdots \mathbf{x}_{[m]})$.  Let $\Lambda_{\mathbf{x}} \subset \mathcal{S}(m,L,w_s)$ be the set of vectors which satisfy the following conditions:
\begin{enumerate}[(a)] 
	\item For every $\mathbf{y} \in \Lambda_{\mathbf{x}}$, exactly $\tilde{m}$ subblocks of $\mathbf{y}$ differ from corresponding subblocks of $\mathbf{x}$.
	\item If $\mathbf{y}_{[j]} \neq \mathbf{x}_{[j]}$, then $d(\mathbf{x}_{[j]}, \mathbf{y}_{[j]}) \in \{1,2\}$.
\end{enumerate}
From the above conditions, it follows that if $\mathbf{y} \in \Lambda_{\mathbf{x}}$, then $d(\mathbf{x},\mathbf{y}) \le 2 \tilde{m} \le t$, and hence $\Lambda_{\mathbf{x}} \subseteq \mathcal{B}_{\mathcal{S}}(\mathbf{x},t;m,L,w_s)$ with
\begin{equation}
|\mathcal{B}_{\mathcal{S}}(\mathbf{x},t;m,L,w_s)| \ge |\Lambda_{\mathbf{x}}| \overset{\text{(i)}}{\ge} \binom{m}{\tilde{m}} \left[(L-w_s)(w_s+1)\right]^{\tilde{m}} ,
\end{equation}
where $\text{(i)}$ follows from Lemma~\ref{Lemma:SECC_subblocks_d12}. Because the above inequality holds for all $\mathbf{x} \in \mathcal{S}(m,L,w_s)$, we have
\begin{equation}
|\mathcal{B}_{\mathcal{S}}^{\mathrm{min}}(t;m,L,w_s)| \ge \binom{m}{\tilde{m}} \left[(L-w_s)(w_s+1)\right]^{\tilde{m}} . \label{eq:SECC_MinBallSizeBound}
\end{equation}
Now,
\begin{equation}
\frac{\tilde{m}}{m} =  \frac{1}{m} \left\lfloor \frac{1}{2} \left\lfloor \frac{\lfloor m L \delta \rfloor -1}{2} \right\rfloor \right\rfloor \implies \lim_{m \to \infty} \frac{\tilde{m}}{m} = \frac{L \delta}{4} \ , \label{eq:tilde_m_limit}
\end{equation}
and hence the claim is proved by combining \eqref{eq:SECC_RateDef}, Prop.~\ref{prop:SECC_SP_CodeSize}, \eqref{eq:SECC_MinBallSizeBound}, and \eqref{eq:tilde_m_limit}.
\end{IEEEproof}
For the case where $L=2$ and $w_s=1$, the asymptotic sphere-packing bound for SECCs reduces to
\begin{equation}
\sigma_{SP}(2,\delta,0.5) = \frac{1}{2}\log 3 - \frac{1}{2} h\left(\frac{\delta}{2}\right) - \frac{\delta}{4} . \label{eq:SECC_SP_rate_L2}
\end{equation}

\section{Rate Penalty Due to Subblock Constraints}\label{sec:penalty}
In this section, we quantify the penalty in rate due to imposition of subblock constraints, relative to the application of corresponding constraints per codeword.
Here, we use the notation $[z]^+$ to imply $\max\{0,z\}$.

\subsection{CWC versus CSCC}
The rate penalty due to constant weight per \emph{subblock}, relative to the constraint requiring constant weight per \emph{codeword}, is quantified by $G_{\alpha - \gamma}(L,\delta,w_s/L)$, defined as
\begin{equation}
G_{\alpha - \gamma}(L,\delta,w_s/L) \triangleq \alpha(\delta,w_s/L) - \gamma(L,\delta,w_s/L) .
\end{equation}
A lower bound to this rate gap is given by
\begin{equation}
G_{\alpha - \gamma}^{LB}(L,\delta,w_s/L) \triangleq \left[\alpha_{GV}(\delta, w_s/L) - \gamma_{SP}(L,\delta,w_s/L)\right]^+ , \label{eq:G_alpha_gamma_def}
\end{equation}
where $\gamma_{SP}(L,\delta,w_s/L)$ is defined in~\eqref{eq:CSCC_SP} and 
\begin{equation}
\alpha_{GV}(\delta,\omega) \triangleq h(\omega) - \omega h\left(\frac{\delta}{2\omega}\right)
- (1-\omega) h\left(\frac{\delta}{2(1-\omega)}\right) \label{eq:CWC_GV},
\end{equation}
with $\alpha_{GV}(\delta, w_s/L)$ denoting the asymptotic GV lower bound for CWCs~\cite{Graham80,McEliece77}. The sphere-packing upper bound on the asymptotic rate for CWCs is given by $\alpha_{SP}(\delta,\omega)$, defined as
\begin{equation}
\alpha_{SP}(\delta,\omega) \triangleq h(\omega) - \omega h\left(\frac{\delta}{4\omega}\right)
- (1-\omega) h\left(\frac{\delta}{4(1-\omega)}\right) \label{eq:CWC_SP}.
\end{equation}

If $L=2$ and $w_s=1$, then using~\eqref{eq:CSCC_Rate_L2} we have the strict inequality $\alpha(\delta,0.5) > \gamma(2,\delta,0.5)$ for $0 < \delta < 0.5$. For relatively large values of the subblock length, $L$, the following theorem shows that rate penalty is strictly positive when $\delta$ is sufficiently small.
\begin{theorem}\label{thm:gap-CWC-CSCC-positive}
	For even $L$ with $L\ge 4$, we have the strict inequality $G_{\alpha - \gamma}^{LB}(L,\delta,0.5) > 0$ for $0 < \delta < \tilde{\delta}_L$, where $\tilde{\delta}_L$ is the smallest positive root of $\tilde{f}_L(\delta)$ defined as
	\begin{equation}
	\tilde{f}_L(\delta) \triangleq 1 - h(\delta) - \frac{1}{L}\log \binom{L}{L/2} + \frac{\delta}{2} \log \frac{L}{2} + \frac{1}{L} h\left(\frac{\delta L}{4}\right) . \label{eq:f_L_tilde_def}
	\end{equation}
\end{theorem}
\begin{IEEEproof}
Using \eqref{eq:CSCC_SP_v2}, \eqref{eq:G_alpha_gamma_def}, and \eqref{eq:CWC_GV}, we have $G_{\alpha - \gamma}^{LB}(L,\delta,0.5) = \tilde{f}_L(\delta)$ when $\delta < 2/L$. We observe from~\eqref{eq:f_L_tilde_def} that $\tilde{f}_L(\delta)$ is a continuous function of $\delta$ with 
\begin{equation}
\tilde{f}_L(0) = 1 - \frac{1}{L}\log \binom{L}{L/2} > 0 . \label{eq:f_L_tilde_positive}
\end{equation} 
Further, when $\delta = 1/L$, we have
\begin{align}
\tilde{f}_L\left(\frac{1}{L}\right) &< 1 - \frac{1}{L}\log \binom{L}{L/2} + \frac{1}{2L} \log \frac{L}{2} + \frac{1}{L} - h\left(\frac{1}{L}\right) \nonumber \\
&\overset{\text{(i)}}{\le} \frac{1}{2L} \log (2L) + \frac{1}{2L} \log \frac{L}{2} + \frac{1}{L} - h\left(\frac{1}{L}\right) \nonumber \\
&= \frac{1}{L} \log (2L) -  h\left(\frac{1}{L}\right) \overset{\text{(ii)}}{\le} 0 , \label{eq:f_L_tilde_negative}
\end{align}
where $\text{(i)}$ and $\text{(ii)}$ follow from~\cite[Ex.~5.8]{GallagerBook68}.
Now using~\eqref{eq:f_L_tilde_positive}, \eqref{eq:f_L_tilde_negative}, and the intermediate value theorem~\cite{RudinBook}, it follows that the equation $\tilde{f}_L(\delta) = 0$ has a solution in the interval $(0,1/L)$. The theorem now follows by denoting the smallest positive root of $\tilde{f}_L(\delta)$ by $\tilde{\delta}_L$.
\end{IEEEproof}
The following proposition addresses the converse question on identifying an interval for $\delta$ when the rate gap between CWCs and CSCCs is provably zero.
\begin{proposition} \label{prop:CWC_CSCC_Rate0}
	The rate gap between CWCs and CSCCs, $G_{\alpha - \gamma}(L,\delta,w_s/L)$, is identically zero when $\delta^* \le \delta \le 1$.
\end{proposition}
\begin{IEEEproof}
	Follows from \eqref{eq:CWC_Rate0} and \eqref{eq:CSCC_Rate0}.	
\end{IEEEproof}

In~\cite{Tandon16_CSCC_TIT}, the gap between  CWC capacity and CSCC capacity on noisy binary input channels was upper bounded by the rate penalty term, $r(L,\omega)$, defined as
\begin{equation}
r(L,\omega) \triangleq h(\omega) - (1/L) \log \binom{L}{L \omega} , \label{eq:CapacityGapUB}
\end{equation}
where $\omega = w_s/L$. Further, it was shown in~\cite{Tandon16_CSCC_TIT} that the actual capacity gap is equal to $r(L,\omega)$ for a noiseless channel. The following proposition shows that $G_{\alpha - \gamma}^{LB}(L,\delta,w_s/L)$ tends to $r(L,w_s/L)$ as $\delta$ tends to 0.
\begin{proposition} \label{prop:CWC_CSCC_gap_eq_rLP}
	For $0 < w_s < L$, we have
	\begin{equation}
	\lim_{\delta \to 0} G_{\alpha - \gamma}^{LB}(L,\delta,w_s/L) = r(L,w_s/L) > 0 .
	\end{equation}
\end{proposition}
\begin{IEEEproof}
	From~\eqref{eq:CWC_GV} we have $\displaystyle \lim_{\delta \to 0} \alpha_{GV}(\delta, w_s/L) = h(w_s/L)$, while using~\eqref{eq:CSCC_SP_v2} we obtain the limit $\displaystyle \lim_{\delta \to 0} \gamma_{SP}(L,\delta,w_s/L) = (1/L) \log \binom{L}{w_s}$, and hence the claim follows from definitions~\eqref{eq:G_alpha_gamma_def} and \eqref{eq:CapacityGapUB}.
\end{IEEEproof}

\begin{proposition}
The lower bound on the rate gap between CWCs and CSCCs, $G_{\alpha - \gamma}^{LB}(L,\delta,w_s/L)$, is tight when $\delta \to 0$.
\end{proposition}
\begin{IEEEproof}
An upper bound on $G_{\alpha - \gamma}(L,\delta,w_s/L)$ is given by $\alpha_{SP}(\delta,w_s/L) - \gamma_{GV}(\delta,\omega)$. Using \eqref{eq:CSCC_GV_L2}, \eqref{eq:CSCC_GV}, and \eqref{eq:CWC_SP}, we observe that this upper bound on the rate gap also tends to $r(L,w_s/L)$ as $\delta$ tends to 0. The proof is complete by combining this observation with Proposition~\ref{prop:CWC_CSCC_gap_eq_rLP}.	
\end{IEEEproof}

\subsection{HWC versus SECC}
In SECCs, the fraction of ones in every subblock is at least $w_s/L$, and hence the fraction of ones in the entire codeword is also at least $w_s/L$. Relative to the constraint requiring at least $w_s/L$ fraction of bits to be 1 for all \emph{codewords}, the rate penalty due to the constraint requiring minimum weight $w_s$ per \emph{subblock} is quantified by $G_{\eta - \sigma}(L,\delta,w_s/L)$, defined as
\begin{equation}
G_{\eta - \sigma}(L,\delta,w_s/L) \triangleq \eta(\delta,w_s/L) - \sigma(L,\delta,w_s/L) .
\end{equation}
For $w_s \ge L/2$, using Theorem \ref{thm:Bachoc}, we note that $G_{\eta - \sigma}(L,\delta,w_s/L)$ can equivalently be expressed as $\alpha(\delta,w_s/L) - \sigma(L,\delta,w_s/L)$ . Thus, a lower bound for $G_{\eta - \sigma}(L,\delta,w_s/L)$, for $w_s \ge L/2$, is given by
\begin{equation}
G_{\eta - \sigma}^{LB}(L,\delta,w_s/L) \triangleq \left[\alpha_{GV}(\delta, w_s/L) - \sigma_{SP}(L,\delta,w_s/L) \right]^+, \label{eq:G_eta_sigma_def}
\end{equation}
where $\alpha_{GV}(\delta, w_s/L)$ and $\sigma_{SP}(L,\delta,w_s/L)$ are defined in~\eqref{eq:CWC_GV} and \eqref{eq:SECC_SP_rate}, respectively. When $w_s \le L/2$, we have $\eta(\delta, w_s/L) = \alpha(\delta, 0.5)$, and in this case, the corresponding rate gap lower bound is defined as 
\begin{equation}
G_{\eta - \sigma}^{LB}(L,\delta,w_s/L) \triangleq \left[\alpha_{GV}(\delta, 0.5) - \sigma_{SP}(L,\delta,w_s/L) \right]^+ .\label{eq:G_eta_sigma_def2}
\end{equation}

The following theorem shows that rate gap between HWCs and SECCs is strictly positive when $\delta$ is sufficiently small.
\begin{theorem}\label{thm:gap-HWC-SECC-positive}
	For even $L$ with $L \ge 4$, we have the strict inequality $G_{\eta - \sigma}^{LB}(L,\delta,0.5) > 0$ for $0 < \delta < \hat{\delta}_L$, where $\hat{\delta}_L$ is the smallest positive root of $\hat{f}_L(\delta)$ defined as
	\begin{equation}
	\hat{f}_L(\delta) \triangleq 1 - h(\delta) - \frac{1}{L} \log \left( \sum_{j=L/2}^L \binom{L}{j} \right) + \frac{1}{L} h\left(\frac{L \delta}{4}\right) + \frac{\delta}{4} \log \left( \frac{L(L+2)}{4} \right) . \label{eq:f_L_hat_def}
	\end{equation}
\end{theorem}
\begin{IEEEproof}
	Using \eqref{eq:G_eta_sigma_def}, \eqref{eq:CWC_GV}, and \eqref{eq:SECC_SP_rate}, we have $G_{\eta - \sigma}^{LB}(L,\delta,0.5) = \hat{f}_L(\delta)$ for $\delta < 2/L$. We observe from~\eqref{eq:f_L_hat_def} that $\hat{f}_L(\delta)$ is a continuous function of $\delta$ with 
	\begin{equation}
	\hat{f}_L(0) = 1 - \frac{1}{L}\log \left( \sum_{j=L/2}^L \binom{L}{j} \right) > 0 . \label{eq:f_L_hat_positive}
	\end{equation} 
	As $\sum_{j=L/2}^{L} \binom{L}{j} > 2^{L-1}$, for $L \ge 4$ we have 
	\begin{align}
	\hat{f}_L\left(\frac{2}{L}\right) &< \frac{2}{L} + \frac{1}{2L} \log \frac{L(L+2)}{4} - h\left(\frac{2}{L}\right) \nonumber \\
	&< \frac{2}{L} + \frac{1}{L} \log (L-1) - h\left(\frac{2}{L}\right) \overset{\text{(i)}}{<} 0 , \label{eq:f_L_hat_negative}
	\end{align}
	where $\text{(i)}$ follow using~\cite[Ex.~5.8]{GallagerBook68}. Now from~\eqref{eq:f_L_hat_positive} and \eqref{eq:f_L_hat_negative}, it follows that the equation $\hat{f}_L(\delta) = 0$ has a solution in the interval $(0,2/L)$. The theorem now follows by denoting the smallest positive root of $\hat{f}_L(\delta)$ by $\hat{\delta}_L$.
\end{IEEEproof}
When $L=2$ and $w_s=1$, it can be verified using~\eqref{eq:SECC_SP_rate_L2} that $G_{\eta - \sigma}^{LB}(2,\delta,0.5) > 0$ for $0 \le \delta < 0.056$.

Theorem~\ref{thm:gap-HWC-SECC-positive} considers the case where $w_s = L/2$. Using a similar argument, it can be shown that in a general setting where $0 < w_s < L$, the rate gap between HWCs and SECCs is strictly positive for sufficiently small $\delta$. The following proposition addresses the converse question on identifying an interval for $\delta$ when this gap is provably zero.

\begin{proposition} \label{prop:HWC_SECC_Rate0}
	For $w_s \le L/2$, the rate gap between HWCs and SECCs, $G_{\eta - \sigma}(L,\delta,w_s/L)$ is identically zero when $1/2 \le \delta \le 1$, while for $w_s \ge L/2$, this gap is zero when $\delta^* \le \delta \le 1$.
\end{proposition}
\begin{IEEEproof}
	The claim for $w_s \le L/2$ follows from \eqref{eq:eta_eq_alpha} and the asymptotic Plotkin bound, while the claim for $w_s \ge L/2$ follows from \eqref{eq:eta_eq_alpha} and \eqref{eq:CWC_Rate0}.
\end{IEEEproof}

\begin{proposition}
The lower bound on the rate gap between HWCs and SECCs, $G_{\eta - \sigma}^{LB}(L,\delta,w_s/L)$, is tight when $\delta \to 0$.
\end{proposition}
\begin{IEEEproof}
For $w_s \le L/2$, from \eqref{eq:G_eta_sigma_def2} we have that
\begin{equation}
G_{\eta - \sigma}^{LB}(L,0,w_s/L) = 1 - \frac{1}{L} \log \left( \sum_{j=w_s}^L \binom{L}{j} \right) . \label{eq:G_eta_sigma_delta0_v1}
\end{equation}
Now, from \eqref{eq:eta_eq_alpha} and the relation $\alpha(\delta, 0.5) = \alpha(\delta)$, an upper bound on $G_{\eta - \sigma}(L,\delta,w_s/L)$ is given by $\alpha_{SP}(\delta, 0.5) - \sigma_{GV}(L,\delta,w_s/L)$. For $w_s \le L/2$, from~\eqref{eq:SECC_GV_rate} and \eqref{eq:CWC_SP}, we note that this upper bound tends to the right hand side of \eqref{eq:G_eta_sigma_delta0_v1} as $\delta \to 0$. This proves the claim for $w_s \le L/2$.

For $w_s \ge L/2$, from \eqref{eq:G_eta_sigma_def} we have that
\begin{equation}
G_{\eta - \sigma}^{LB}(L,0,w_s/L) = h(w_s/L) - \frac{1}{L} \log \left( \sum_{j=w_s}^L \binom{L}{j} \right) . \label{eq:G_eta_sigma_delta0_v2}
\end{equation}
For $w_s \ge L/2$, an upper bound on the rate gap $G_{\eta - \sigma}(L,\delta,w_s/L)$ is given by $\alpha_{SP}(\delta, w_s/L) - \sigma_{GV}(L,\delta,w_s/L)$ (using \eqref{eq:eta_eq_alpha}), and this upper bound tends to the right hand side of \eqref{eq:G_eta_sigma_delta0_v2} as $\delta \to 0$.

\end{IEEEproof}

\subsection{SECC versus CSCC}
The SECCs, relative to CSCCs, provide the flexibility of allowing different subblocks to have different weights. In this subsection, we show that this flexibility leads to an improvement in asymptotic rate when the relative distance of the code is sufficiently small. The gap between SECC rate and CSCC rate is quantified by $G_{\sigma - \gamma}(L,\delta,w_s/L)$, defined as
\begin{equation}
G_{\sigma - \gamma}(L,\delta,w_s/L) \triangleq \sigma(L,\delta,w_s/L) - \gamma(L,\delta,w_s/L) .
\end{equation}
A lower bounded to this rate gap is given by
\begin{equation}
G_{\sigma - \gamma}^{LB}(L,\delta,w_s/L) \triangleq \left[\sigma_{GV}(L,\delta, w_s/L) - \gamma_{SP}(L,\delta,w_s/L)\right]^+ , \label{eq:G_sigma_gamma_def}
\end{equation}
where $\sigma_{GV}(L,\delta,w_s/L)$ and $\gamma_{SP}(L,\delta,w_s/L)$ are given by \eqref{eq:SECC_GV_rate} and \eqref{eq:CSCC_SP}, respectively. The following theorem shows that $G_{\sigma - \gamma}^{LB}(L,\delta,w_s/L)$ is strictly positive when $\delta$ is small.

\begin{theorem}\label{thm:gap-SECC-CSCC}
	For even $L$ with $L\ge 4$, we have the strict inequality $G_{\sigma - \gamma}^{LB}(L,\delta,0.5) > 0$ for $0 < \delta < \grave{\delta}_L$, where $\grave{\delta}_L$ is the smallest positive root of $\grave{f}_L(\delta)$ defined as
	\begin{align}
	\grave{f}_L(\delta) \triangleq \frac{1}{L} \log \left( \sum_{j=L/2}^L \binom{L}{j} \right) - h(\delta) - \frac{1}{L}\log \binom{L}{L/2} + \frac{\delta}{2} \log \frac{L}{2} + \frac{1}{L} h\left(\frac{\delta L}{4}\right) . \label{eq:f_L_grave_def}
	\end{align}
\end{theorem}
\begin{IEEEproof}
Using \eqref{eq:CSCC_SP_v2}, \eqref{eq:SECC_GV_rate}, and \eqref{eq:G_sigma_gamma_def}, we have $G_{\sigma - \gamma}^{LB}(L,\delta,0.5) = \grave{f}_L(\delta)$ for $\delta < 2/L$. From \eqref{eq:f_L_grave_def} we note that $\grave{f}_L(\delta)$ is a continuous function of $\delta$ with
\begin{equation}
\grave{f}_L(0) = \frac{1}{L} \log \left( \sum_{j=L/2}^L \binom{L}{j} \right) - \frac{1}{L} \log \binom{L}{L/2} > 0 . \label{eq:f_L_grave_positive}
\end{equation}
Further, comparing \eqref{eq:f_L_tilde_def} and \eqref{eq:f_L_grave_def}, we observe that $\grave{f}_L(\delta) < \tilde{f}_L(\delta)$, In particular, for $\delta = 1/L$ we have
\begin{equation}
\grave{f}_L\left(\frac{1}{L} \right) < \tilde{f}_L\left(\frac{1}{L} \right) < 0 , \label{eq:f_L_grave_negative}
\end{equation}
where the last inequality follows from \eqref{eq:f_L_tilde_negative}. From \eqref{eq:f_L_grave_positive} and \eqref{eq:f_L_grave_negative} it follows that the equation $\grave{f}_L(\delta) = 0$ has a solution in the interval $(0,1/L)$. The proof is complete be denoting the smallest positive root of $\grave{f}_L(\delta)$ by $\grave{\delta}_L$.
\end{IEEEproof}
For the case when $L=2$ and $w_s=1$, we have
\begin{equation}
G_{\sigma - \gamma}^{LB}(2,\delta,0.5) = \left[ \frac{1}{2}\log (3/2) - h(\delta) + \frac{1}{2}h(\delta/2) \right]^+ , \label{eq:G_sigma_gamma_L2}
\end{equation}
and $G_{\sigma - \gamma}^{LB}(2,\delta,0.5)$ is strictly positive for $0 \le \delta < 0.084$.

From Proposition~\ref{prop:SECC_GV_rate} and Theorem~\ref{thm:SECC_SP_rate}, note that for $0 < \delta < \min\{\delta^*, 4/L\}$, we have $\sigma_{GV}(L,\delta,w_s/L) \le \sigma_{SP}(L,\delta,w_s/L)$, and hence it follows from definitions~\eqref{eq:G_alpha_gamma_def}, \eqref{eq:G_eta_sigma_def}, and \eqref{eq:G_sigma_gamma_def}  that
\begin{equation}
G_{\alpha - \gamma}^{LB}(L,\delta,w_s/L) \ge G_{\alpha - \sigma}^{LB}(L,\delta,w_s/L) + G_{\sigma - \gamma}^{LB}(L,\delta,w_s/L) . \label{eq:G_inequality}
\end{equation}

Although Theorem~\ref{thm:gap-SECC-CSCC} only considers the case $w_s = L/2$, a similar argument can be applied to show that the rate gap between SECCs and CSCCs is strictly positive in a general setting where $0 < w_s < L$, provided $\delta$ is sufficiently small. The following converse, providing an interval for $\delta$ which results in zero rate gap, is obtained by using an argument similar to that in Proposition~\ref{prop:HWC_SECC_Rate0}.
\begin{proposition} \label{prop:SECC_CSCC_Rate0}
	For $w_s \le L/2$, the rate gap between SECCs and CSCCs, $G_{\sigma - \gamma}(L,\delta,w_s/L)$, is identically zero when $1/2 \le \delta \le 1$, while for $w_s \ge L/2$, this gap is zero when $\delta^* \le \delta \le 1$.
\end{proposition}

The following proposition establishes the tightness of $G_{\sigma - \gamma}^{LB}(L,\delta,w_s/L)$ when $\delta$ tends to 0.
\begin{proposition}
	The lower bound on the rate gap between SECCs and CSCCs, $G_{\sigma - \gamma}^{LB}(L,\delta,w_s/L)$, is tight when $\delta \to 0$.
\end{proposition}
\begin{IEEEproof}
From~\eqref{eq:CSCC_SP} and \eqref{eq:SECC_GV_rate}, we have that
\begin{equation}
G_{\sigma - \gamma}^{LB}(L,0,w_s/L) = \frac{1}{L} \log \left( \sum_{j=w_s}^L \binom{L}{j} \right)  - \frac{1}{L} \log \binom{L}{w_s} . \label{eq:G_sigma_gamma_delta0}
\end{equation}
An upper bound on $G_{\sigma - \gamma}(L,\delta,w_s/L)$ is given by $\sigma_{SP}(L,\delta,w_s/L) - \gamma_{GV}(L,\delta,w_s/L)$. From~\eqref{eq:CSCC_GV_L2}, \eqref{eq:CSCC_GV}, and \eqref{eq:SECC_SP_rate}, we note that this upper bound tends to the right hand side of~\eqref{eq:G_sigma_gamma_delta0} as $\delta \to 0$. 
\end{IEEEproof}

\section{Numerical Results}
\label{sec:numerical}

In this section, we provide numerical bounds on rate penalties due to weight constraint per subblock, relative to imposing similar constraint per codeword. 

\begin{figure}[t]
\centering \includegraphics[width=0.55\textwidth]{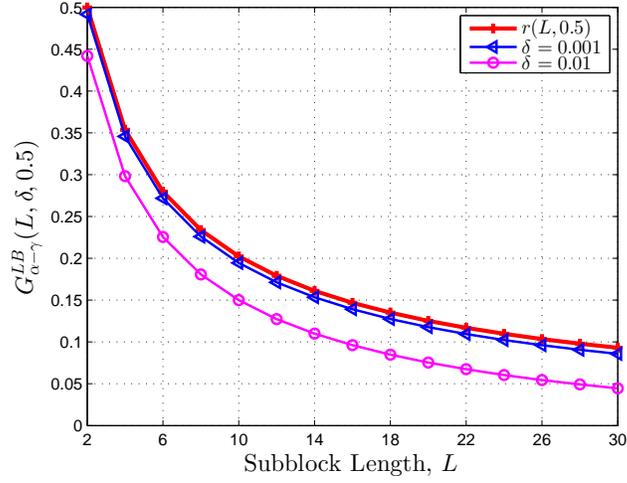}
\centering \caption{$G_{\alpha - \gamma}^{LB}(L,\delta,0.5)$ versus subblock length, $L$.}
\label{Fig:G_CWC_CSCC_versus_L}
\end{figure}

\begin{figure}[t]
	\centering
	\includegraphics[width=0.55\textwidth]{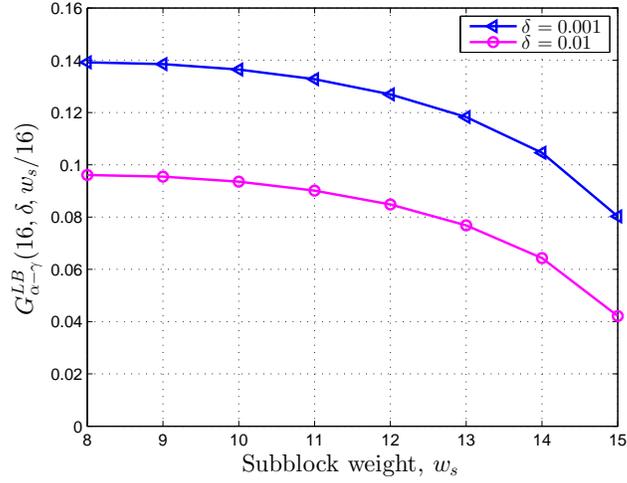}
	\caption{$G_{\alpha - \gamma}^{LB}(16,\delta,w_s/16)$ as a function of subblock weight, $w_s$.}
	\label{Fig:G_CWC_CSCC_versus_omega}
\end{figure}

\begin{figure}[t]
	\centering
	\includegraphics[width=0.55\textwidth]{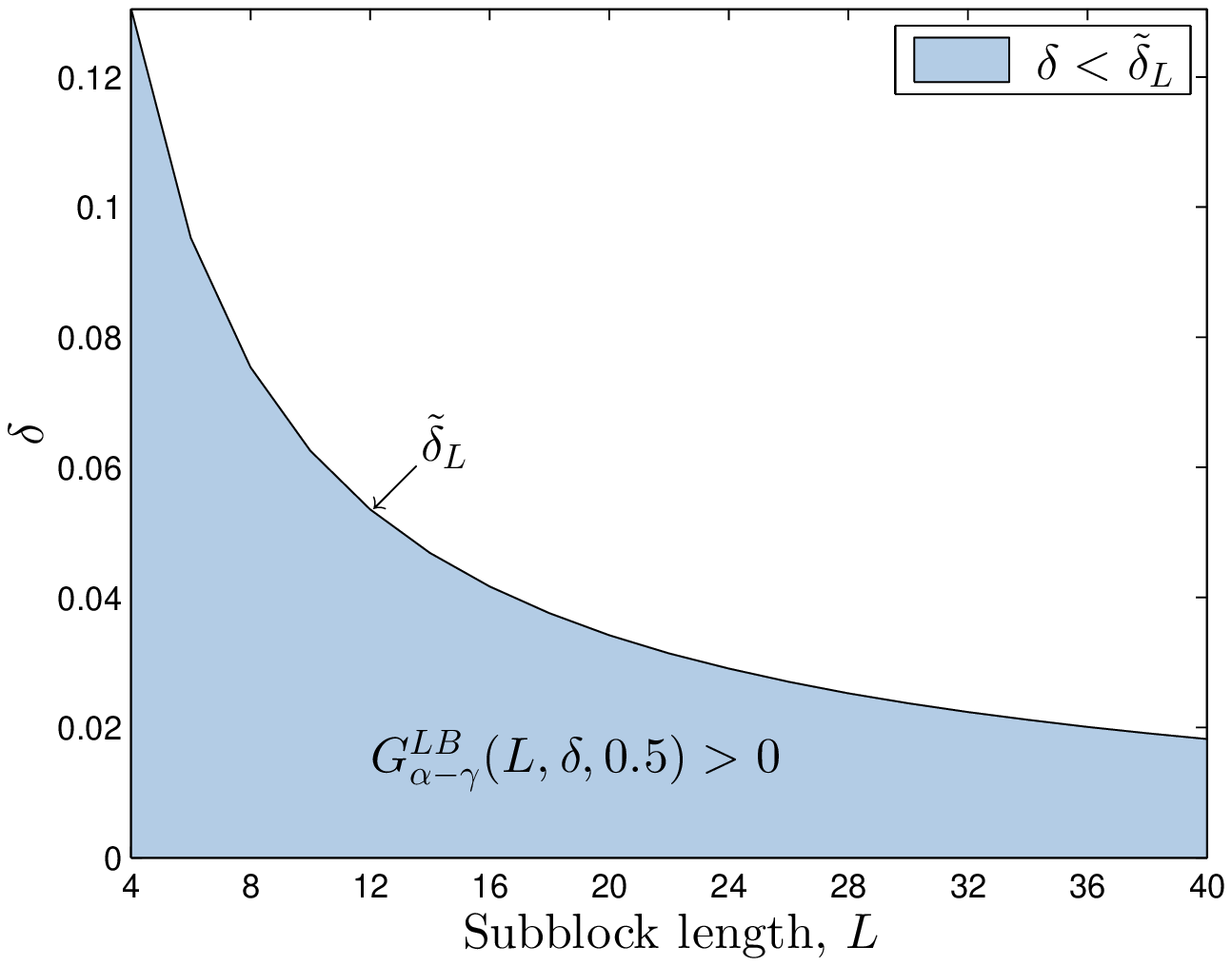}
	\caption{Area where $G_{\alpha - \gamma}^{LB}(L,\delta,0.5)$ is strictly positive.}
	\label{Fig:delta_L_tilde}
\end{figure}

Fig.~\ref{Fig:G_CWC_CSCC_versus_L} plots the lower bound on the rate gap between CWCs and CSCCs, $G_{\alpha - \gamma}^{LB}(L,\delta,0.5)$, as a function of the subblock length. The upper bound on the gap between CWC capacity and CSCC capacity on noisy binary input channels for $w_s = L/2$, given by $r(L,0.5)$ (see~\eqref{eq:CapacityGapUB}), is also plotted in red. As suggested by Proposition~\ref{prop:CWC_CSCC_gap_eq_rLP}, the figure shows that $G_{\alpha - \gamma}^{LB}(L,\delta,0.5)$ tends to $r(L,0.5)$ as $\delta$ gets close to zero. For a fixed value of $w_s/L$, note that $\alpha_{GV}(\delta, w_s/L)$ is independent of $L$. Thus, for a given $\delta$, the decrease in $G_{\alpha - \gamma}^{LB}(L,\delta,0.5)$ with increasing $L$ is due to an increase in CSCC rate. This is intuitively expected, because an increase in $L$ allows for greater flexibility in the choice of bits within every subblock. Further, from Proposition~\ref{prop:Chee}, it follows that $G_{\alpha - \gamma}^{LB}(L,\delta,0.5) \to 0$ as $L \to \infty$.

Fig.~\ref{Fig:G_CWC_CSCC_versus_omega} plots $G_{\alpha - \gamma}^{LB}(L,\delta,w_s/L)$ when the subblock length is fixed at $L=16$, and $w_s$ varies from $L/2 = 8$ to $L-1 = 15$. Note that $\alpha_{GV}(L,\delta, (L-w_s)/L) = \alpha_{GV}(L,\delta, w_s/L)$ and $\gamma_{SP}(L,\delta,(L-w_s)/L) = \gamma_{SP}(L,\delta,w_s/L)$ (see~\eqref{eq:CSCC_SP} and \eqref{eq:CWC_GV}, respectively), and thus $G_{\alpha - \gamma}^{LB}(L,\delta,(L-w_s)/L) = G_{\alpha - \gamma}^{LB}(L,\delta,w_s/L)$. Note that Figs.~\ref{Fig:G_CWC_CSCC_versus_L} and \ref{Fig:G_CWC_CSCC_versus_omega} illustrate that $G_{\alpha - \gamma}^{LB}(L,\delta,w_s/L)$ decreases with $\delta$.

Fig.~\ref{Fig:delta_L_tilde} depicts the region where the gap between CWC rate and CSCC rate is provably strictly positive. Note that $\tilde{\delta}_L$ is the smallest value of $\delta$ for which the lower bound $G_{\alpha - \gamma}^{LB}(L,\delta,w_s/L)$ is zero, when $L$ is fixed, and $w_s=L/2$ (see Theorem~\ref{thm:gap-CWC-CSCC-positive}). The figure shows that $\tilde{\delta}_L$ decreases with $L$, and from Proposition~\ref{prop:Chee} it follows that $\tilde{\delta}_L \to 0$ when $L \to \infty$. Moreover, using Proposition~\ref{prop:CWC_CSCC_Rate0}, it is seen that the actual rate gap $G_{\alpha - \gamma}(L,\delta,0.5)$ is provably zero for $\delta \ge 0.5$.

\begin{figure}[t]
	\centering
	\includegraphics[width=0.55\textwidth]{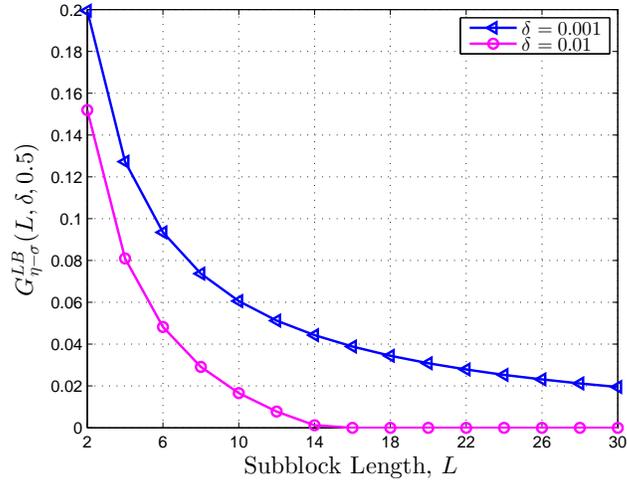}
	\caption{$G_{\eta - \sigma}^{LB}(L,\delta,0.5)$ as a function of subblock length, $L$.}
	\label{Fig:G_HWC_SECC_versus_L}
\end{figure}

\begin{figure}[t]
	\centering
	\includegraphics[width=0.55\textwidth]{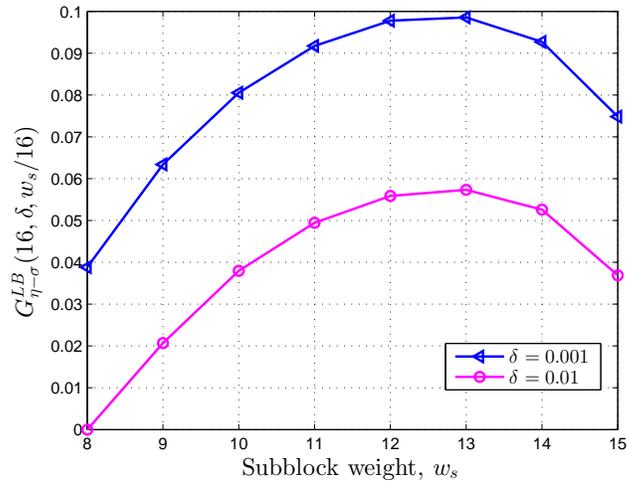}
	\caption{$G_{\eta - \sigma}^{LB}(16,\delta,w_s/16)$ as a function of $w_s$.}
	\label{Fig:G_HWC_SECC_versus_omega}
\end{figure}

\begin{figure}[t]
	\centering
	\includegraphics[width=0.55\textwidth]{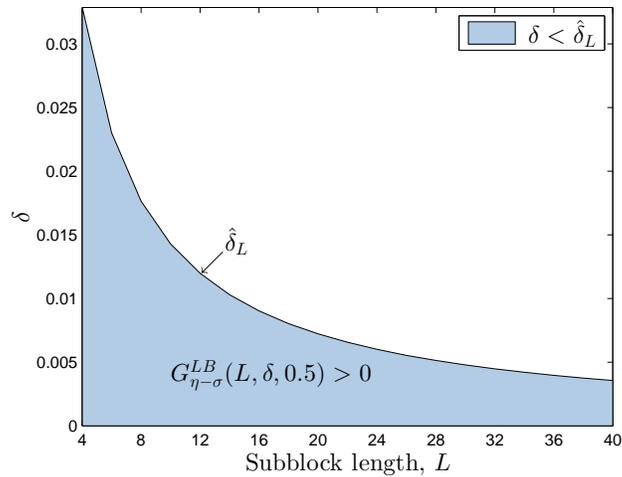}
	\caption{Area where $G_{\eta - \sigma}^{LB}(L,\delta,0.5)$ is strictly positive.}
	\label{Fig:delta_L_hat}
\end{figure}

Fig.~\ref{Fig:G_HWC_SECC_versus_L} plots $G_{\eta - \sigma}^{LB}(L,\delta,w_s/L)$, lower bound for the rate gap between HWCs and SECCs, as a function of $L$, with $w_s=L/2$. For a given $\delta$, it is seen from the figure that $G_{\eta - \sigma}^{LB}(L,\delta,0.5)$ decreases with $L$. Note that for $w_s \ge L/2$, using Proposition~\ref{prop:BachocChee}, we have $G_{\eta - \sigma}^{LB}(L,\delta,w_s/L) \to 0$ as $L~\to~\infty$. Fig.~\ref{Fig:G_HWC_SECC_versus_omega} plots $G_{\eta - \sigma}^{LB}(L,\delta,w_s/L)$ versus $w_s$, for fixed $L=16$.

The shaded area in Fig.~\ref{Fig:delta_L_hat} depicts the region where the rate gap between HWC and SECC is provably strictly positive. Here, $\hat{\delta}_L$ is the smallest value of $\delta$ for which the lower bound $G_{\eta - \sigma}^{LB}(L,\delta,w_s/L)$ is zero, when $L$ is fixed, and $w_s=L/2$ (see Theorem~\ref{thm:gap-HWC-SECC-positive}). The figure shows that $\hat{\delta}_L$ decreases with $L$, and from Proposition~\ref{prop:BachocChee} it follows that $\hat{\delta}_L \to 0$ when $L \to \infty$. Moreover, using Proposition~\ref{prop:HWC_SECC_Rate0}, it is seen that the actual rate gap $G_{\eta - \sigma}(L,\delta,0.5)$ is provably zero for $\delta \ge 0.5$.

\begin{figure}[t]
	\centering
	\includegraphics[width=0.55\textwidth]{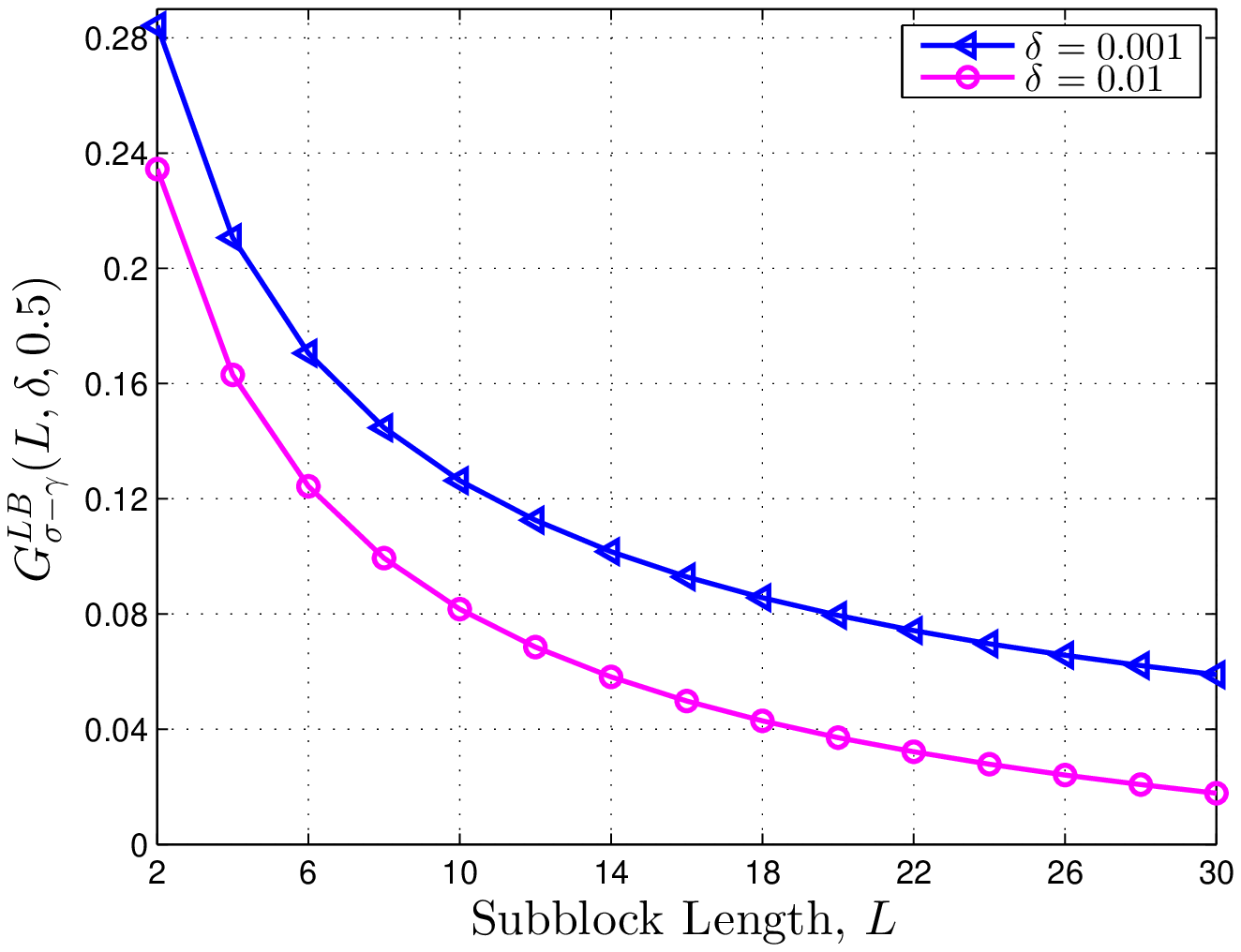}
	\caption{$G_{\sigma - \gamma}^{LB}(L,\delta,0.5)$ versus subblock length, $L$.}
	\label{Fig:G_SECC_CSCC_versus_L}
\end{figure}

\begin{figure}[t]
	\centering
	\includegraphics[width=0.55\textwidth]{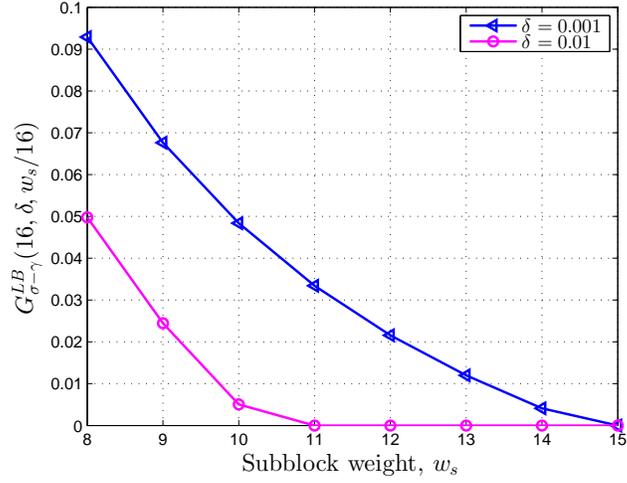}
	\caption{$G_{\sigma - \gamma}^{LB}(16,\delta,w_s/16)$ as a function of $w_s$.}
	\label{Fig:G_SECC_CSCC_versus_omega}
\end{figure}

\begin{figure}[t]
	\centering
	\includegraphics[width=0.55\textwidth]{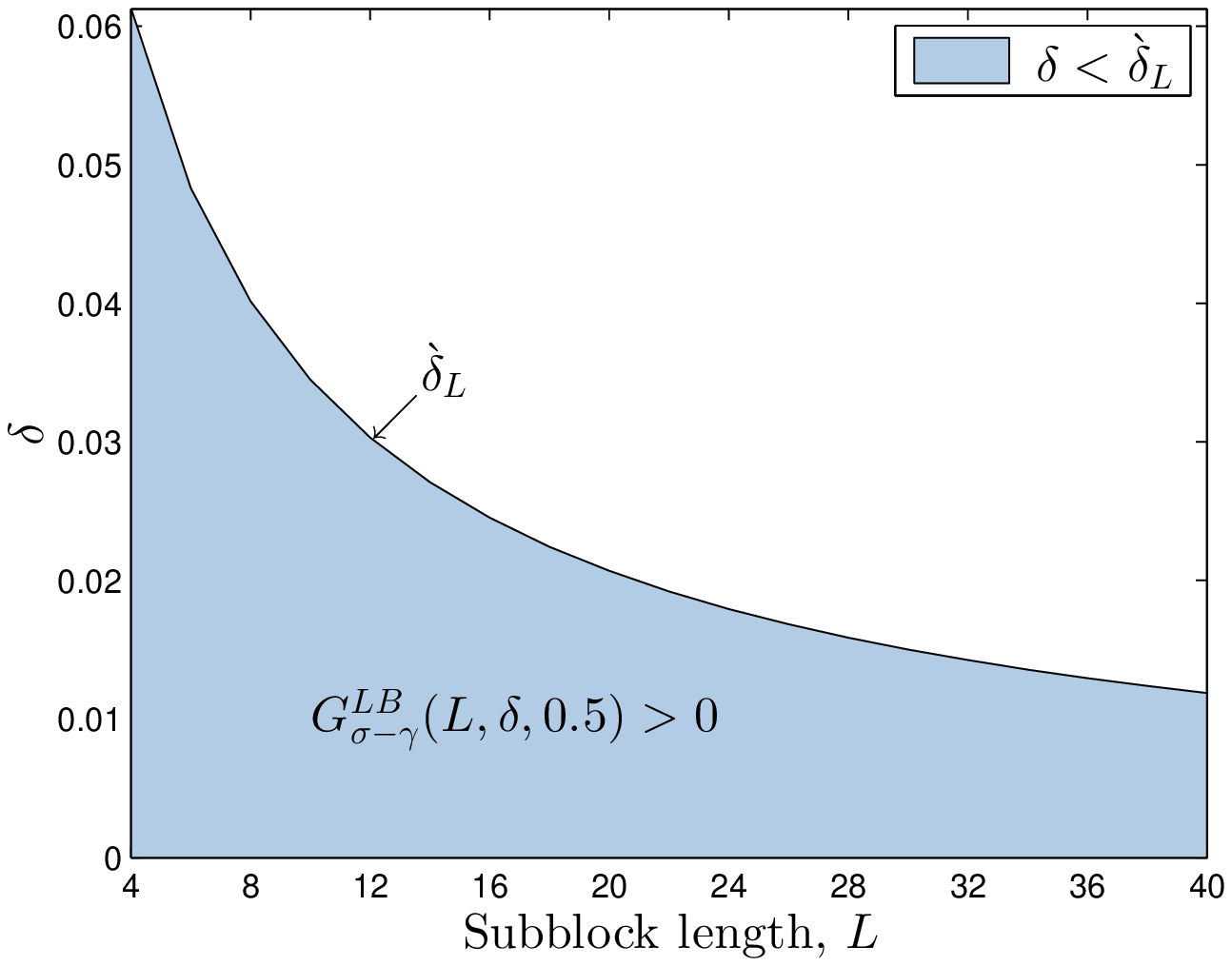}
	\caption{Area where $G_{\sigma - \gamma}^{LB}(L,\delta,0.5)$ is strictly positive.}
	\label{Fig:delta_L_grave}
\end{figure}

Relative to CSCCs, the SECCs allow for greater flexibility in choice of bits within each subblock, by allowing the subblock weight to vary, provided it exceeds a certain threshold. This flexibility results in higher rate for SECCs and Fig.~\ref{Fig:G_SECC_CSCC_versus_L} plots $G_{\sigma - \gamma}^{LB}(L,\delta,0.5)$, lower bound on the rate gap between SECCs and CSCCs. The figure shows that for a given $\delta$, the rate gap bound decreases with $L$, and we have $G_{\sigma - \gamma}^{LB}(L,\delta,0.5) \to 0$ as $L \to \infty$. The last assertion follows by combining Theorem~\ref{thm:Bachoc}, Proposition~\ref{prop:Chee}, and the fact that $\gamma(L,\delta,w_s/L) \le \sigma(L,\delta,w_s/L) \le \eta(\delta,w_s/L)$. Additionally, comparing Figs.~\ref{Fig:G_CWC_CSCC_versus_L}, \ref{Fig:G_HWC_SECC_versus_L}, and \ref{Fig:G_SECC_CSCC_versus_L}, we observe that the inequality in \eqref{eq:G_inequality} is satisfied.

Fig.~\ref{Fig:G_SECC_CSCC_versus_omega} plots $G_{\sigma - \gamma}^{LB}(L,\delta,w_s/L)$ versus $w_s$, for fixed $L=16$, and $\delta \in \{0.001, 0.01\}$. On comparing Figs.~\ref{Fig:G_CWC_CSCC_versus_omega}, \ref{Fig:G_HWC_SECC_versus_omega}, and \ref{Fig:G_SECC_CSCC_versus_omega}, it is observed that lower bounds on respective rate gaps satisfy \eqref{eq:G_inequality}. Fig.~\ref{Fig:delta_L_grave} depicts the region where the rate gap between SECC and CSCC is provably strictly positive. Here, $\grave{\delta}_L$ is the smallest value of $\delta$ for which the lower bound $G_{\sigma - \gamma}^{LB}(L,\delta,w_s/L)$ is zero, when $L$ is fixed, and $w_s=L/2$ (see Theorem~\ref{thm:gap-SECC-CSCC}). Fig.~\ref{Fig:delta_L_grave} shows that $\grave{\delta}_L$ decreases with $L$, and from Thm.~\ref{thm:Bachoc}, Prop.~\ref{prop:Chee}, and Prop.~\ref{prop:BachocChee} it follows that $\grave{\delta}_L \to 0$ when $L \to \infty$. Moreover, using Proposition~\ref{prop:SECC_CSCC_Rate0}, we see that the true gap $G_{\sigma - \gamma}(L,\delta,0.5)$ is provably zero for $\delta \ge 0.5$.

\section{Concluding Remarks}

\balance

We derived upper and lower bounds for the sizes of CSCCs and SECCs.
For a fixed subblock length $L$ and weight parameter $w_s$, 
we demonstrated the existence of some $\tilde{\delta}_L$, $\hat{\delta}_L$, and $\grave{\delta}_L$ such that the gaps 
\begin{align*}
G_{\alpha - \gamma}(L,\delta,w_s/L) &>0& \mbox{ for } \delta < \tilde{\delta}_L,\\
G_{\eta - \sigma}(L,\delta,w_s/L)&>0& \mbox{ for } \delta < \hat{\delta}_L,\mbox{ and,}\\
G_{\sigma - \gamma}(L,\delta,w_s/L)&>0& \mbox{ for } \delta < \grave{\delta}_L.
\end{align*}
Furthermore, we provide estimates on $\tilde{\delta}_L$, $\hat{\delta}_L$, and $\grave{\delta}_L$
via Theorems \ref{thm:gap-CWC-CSCC-positive}, \ref{thm:gap-HWC-SECC-positive}, and \ref{thm:gap-SECC-CSCC}.
These gaps then reflect the rate penalties due to imposition of subblock constraints, relative to the application of corresponding constraints per codeword. 


The converse problem, on identifying an interval for $\delta$ where the respective rate penalties are provably zero, is addressed via Propositions~\ref{prop:CWC_CSCC_Rate0}, \ref{prop:HWC_SECC_Rate0}, and \ref{prop:SECC_CSCC_Rate0}. An interesting but unsolved problem in this regard is to characterize the smallest $\delta$ beyond which the respective rate penalties are zero.  We can get some insight from the numerical computations in \cite{Tandon16_CSCC_TIT}, which indicate that there is a nonzero gap between CSCC and CWC capacities and a nonzero gap between CSCC and SECC capacities.  This suggests that, for a fixed subblock length $L$, the rate penalties are zero if and only if the respective asymptotic rates themselves are zero.  However, this remains an open problem.


\end{document}